\documentclass[journal=jctcce,manuscript=article]{achemso}
\usepackage[utf8]{inputenc}
\usepackage[T1]{fontenc}
\usepackage{siunitx}
\usepackage{graphicx}
\usepackage{hyperref}
\usepackage{amsmath}
\usepackage{amssymb}
\usepackage[version=4]{mhchem}
\usepackage[inline]{enumitem}
\usepackage{braket}
\usepackage{float}
\usepackage{multirow}
\usepackage{makecell}
\usepackage{ifthen}
\usepackage{booktabs}

\usepackage{xspace}
\makeatletter
\def\ie{\textit{i.e.}\@\xspace}
\def\eg{\textit{e.g.}\@\xspace}
\makeatother
\newcommand{\SPAHMinner}{SPA\textsuperscript{H}M}
\newcommand\SPAHM[1][9]{\SPAHMinner\ifthenelse{\equal{#1}{9}}{\xspace}{(#1)}}
\newcommand{\eSPAHM}{$\varepsilon$-\SPAHM}
\renewcommand{\vec}[1]{\mathrm{\mathbf{#1}}}
\newcommand{\de}{\,{\mathrm{d}}}
\newcommand{\transp}{^\intercal}
\newcommand{\phantomtransp}{^{\vphantom{\intercal}}}
\newcommand{\kovlp}[1]{K^\mathrm{overlap}_{#1}}

\graphicspath{{fig/}}

\newcommand{\extref}[1]{\ref{S-#1}}
\title{\texorpdfstring{\SPAHM[a,b]}{SPAHM(a,b)}: encoding the density information from guess Hamiltonian in quantum machine learning representations} 

\author{Ksenia R. Briling}
\affiliation{Laboratory for Computational Molecular Design, Institute of Chemical Sciences and Engineering,
\'{E}cole Polytechnique F\'{e}d\'{e}rale de Lausanne, 1015 Lausanne, Switzerland}
\altaffiliation{These authors contributed equally to this work.}
\author{Yannick Calvino Alonso}
\affiliation{Laboratory for Computational Molecular Design, Institute of Chemical Sciences and Engineering,
\'{E}cole Polytechnique F\'{e}d\'{e}rale de Lausanne, 1015 Lausanne, Switzerland}
\altaffiliation{These authors contributed equally to this work.}
\author{Alberto Fabrizio}
\affiliation{Laboratory for Computational Molecular Design, Institute of Chemical Sciences and Engineering,
\'{E}cole Polytechnique F\'{e}d\'{e}rale de Lausanne, 1015 Lausanne, Switzerland}
\alsoaffiliation{National Centre for Computational Design and Discovery of Novel Materials (MARVEL),
\'{E}cole Polytechnique F\'{e}d\'{e}rale de Lausanne, 1015 Lausanne, Switzerland}
\author{Clemence Corminboeuf}
\email{clemence.corminboeuf@epfl.ch}
\affiliation{Laboratory for Computational Molecular Design, Institute of Chemical Sciences and Engineering,
\'{E}cole Polytechnique F\'{e}d\'{e}rale de Lausanne, 1015 Lausanne, Switzerland}
\alsoaffiliation{National Centre for Computational Design and Discovery of Novel Materials (MARVEL),
\'{E}cole Polytechnique F\'{e}d\'{e}rale de Lausanne, 1015 Lausanne, Switzerland}

\begin{document}

\begin{abstract}
Recently, we introduced a class of molecular representations for kernel-based regression methods ---
the spectrum of approximated Hamiltonian matrices (\SPAHM) --- that takes advantage of
lightweight one-electron Hamiltonians traditionally used as an SCF initial guess.
The original \SPAHM variant is built from occupied-orbital energies (\ie, eigenvalues)
and naturally contains all the information about nuclear charges, atomic positions, and symmetry requirements.
Its advantages were demonstrated on datasets featuring a wide variation of charge and spin,
for which traditional structure-based representations commonly fail.
\SPAHM[a,b], as introduced here, expand the eigenvalue \SPAHM into local and transferable representations.
They rely upon one-electron density matrices to build fingerprints
from atomic and bond density overlap contributions
inspired from preceding state-of-the-art representations.
The performance and efficiency of \SPAHM[a,b] is assessed on the predictions for datasets of
prototypical organic molecules (QM7) of different charges
and azoheteroarene dyes in an excited state.
Overall, both \SPAHM[a] and \SPAHM[b]
outperform state-of-the-art representations
on difficult prediction tasks such as the
atomic properties of charged open-shell species and of $\pi$-conjugated systems.
\end{abstract}

\section{Introduction}

Physics-based machine learning representations, also known as representations for quantum machine learning (QML),
\cite{FMFC2019,MGBOCC2021,KVCCGMT2021,HL2021,LGP2022}
are rooted in the fundamental principle that all the (static) information about a neutral chemical
system is uniquely encoded into the system-specific parameters that fix the electronic
Schr\"{o}dinger equation: nuclear charges $\{Z_I\}$ and positions $\{\vec R_I\}$. Owing to their physical
origins, these representations are highly general and have a deep connection to quantum-chemical
targets. Hence, they have been broadly exploited to supply fast and accurate predictions of a myriad
of atomistic chemical properties.

To ensure efficient predictions, most QML representations encode the information relative to the
atoms and their environment through the derivation of rather simple non-linear functions
of $\{Z_I\}$ and $\{\vec R_I\}$ thus bypassing the construction of the Hamiltonian entirely.
Most popular examples include representations built from internal coordinates (MBTR,\cite{HR2017}
PIPs,\cite{BMBJB2004,BB2009,BBCCCFHKSSWX2010,XB2010,JG2013}
and graph-based representations\cite{PA2011});
those that encode regions of atomic geometries by using a local expansion of a Gaussian smeared atomic density
(Behler--Parrinello symmetry functions,\cite{BP2007,B2011,ZHWCE2018}
smooth overlap of atomic positions (SOAP),\cite{BKC2013,GWCC2018}
the overlap fingerprint,\cite{ZAFSFRGSGWG2016}
NICE,\cite{NPC2020}
and ACE\cite{D2019a,D2019b,DBCDEOO2022});
as well as those based on values or fingerprints of physics-inspired potentials
(Coulomb matrix,\cite{RTML2012,RRL2015}
bag of bonds,\cite{HBRPLMT2015}
(a)SLATM,\cite{HL2020}
LODE,\cite{GC2019}
FCHL18,\cite{FCHL2018}
and FCHL19\cite{CBFL2020}).

Each of these categories of representations have led to impressive performances for the predictions
of both prototypical and complex molecular or material properties\cite{FHHGSDVKRL2017}
such as
atomization energies,\cite{RTML2012}
multipole moments,\cite{BAV2015}
polarizabilities,\cite{GWCC2018,WGYLDC2019}
HOMO--LUMO gaps,\cite{MRGVHTML2013,MSL2022}
molecular forces,\cite{LKD2015,CTSPSM2017,CSMT2018}
potential energy surfaces,\cite{Behler2017,BP2007,SNLIR2018}
electron densities,\cite{BVLTBM2017,GFMWCC2019,FGMCC2019,CKBKCR2019}
density functionals,\cite{ZHWC2004}
and many-body wavefunctions\cite{SGTMM2019}.
Yet, since such representations are functions of $\{Z_I\}$ and $\{\vec R_I\}$ only,
achieving the same level of accuracy for chemical
targets inherently dependent upon changes in electron delocalization, spin, or charge remains a challenge
and additional electronic information (\ie, the Hamiltonian) is needed.
An alternative approach consists in adding
one more layer between the geometry and the representation
and complementing the latter with some quantum-chemical information
computed from the former.
Illustrative examples include OrbNet,\cite{QWAMM2020,CSQOSDBAWMM2021}
which uses quantum-mechanical operators
obtained from a converged semiempirical computation
as input features for a neural network, as well as methodologies such as EHML-ML~\cite{LCTGY2018}
and DFTB-ML~\cite{ZNLSZZKBIT2021} aiming at refining the parameters
characteristic of semiempirical methods (\eg, H\"{u}ckel theory and DFTB) to achieve higher-level accuracy.
Alternative models like EPNN\cite{MJSCS2021} propose a heuristic neural-network-based
partitioning scheme
to provide fast and reliable quantum-like atomic charges as input for predictive models.
AIMNet\cite{ZSLI2019} with the neural spin-charge equilibration unit\cite{ZSNTI2021}
takes $\{\vec R_I\}$, $\{Z_I\}$, and total molecular charge and spin multiplicity
to learn a state-specific representation with a message-passing neural network.
More computationally demanding alternatives consist in featurizing components of fully converged Hartree--Fock-level matrices,
operators, densities, or determinants, as in DeePHF~\cite{CZWE2020}, DeePKS~\cite{CZWE2020b},
MO-ML~\cite{WCM2018,CWCM2019,CSM2022},
the orbital-based FJK representation,\cite{KL2022} and the kernel density functional approximation\cite{MR2021} (KDFA).
Also relevant to this category is the recent introduction\cite{GR2022}
of Coulomb lists and smooth overlap of electron densities that bridge
geometry-based descriptors with electronic structure theory.
The recently introduced matrix of orthogonalized atomic orbital coefficients proposes a compact
although more expensive representation derived from an orbital localization scheme.\cite{LG2023}

With the same purpose of encoding valuable electronic information,
we recently introduced the spectrum of approximated Hamiltonian matrices (\SPAHM) representation family,\cite{FBC2022}
which has the advantage of avoiding the self-consistent field (SCF) procedure.
Specifically, the eigenvalue \SPAHM (\eSPAHM) is a compact global representation
consisting of occupied-orbital eigenvalues extracted from lightweight one-electron Hamiltonians
traditionally used as an SCF initial guess in molecular quantum chemistry codes.

Owing to a seamless generalization to open-shell systems, \eSPAHM
performs well on datasets characterized by a wide variation of charge and spin,
for which the traditional structure-based representations commonly fail.
However, it suffers from some limitations:
\begin{enumerate*}[label=\roman*),itemjoin={{ }}, itemjoin*={{ and }}]
\item
its global nature limits transferability,\cite{CGBMDC2023}
\item
it only exploits eigenvalues, despite the availability of additional information
(\eg, the eigenvectors and associated electron densities),
\item
comparing the orbital energies of compounds having different size and composition lacks physical sense
\end{enumerate*}.

To address such limitations, in this work we expand \SPAHM and build two types of representations
exploiting the electron density extracted from the same approximated Hamiltonians.
We then bridge the conceptual advantages of both SOAP\cite{BKC2013}
and atomic version of SLATM\cite{HL2020} (aSLATM)
to obtain atomic-density overlap fingerprints, \SPAHM[a], or bond-density based representation, \SPAHM[b].

The predictive power of \SPAHM[a,b] is demonstrated on local (atomic) properties
such as atomic partial charges, spin densities, and isotropic magnetic shielding
on the QM7 dataset.\cite{BR2009,RTML2012}
We then show the excellent performance of the models
on datasets made of a mix of neutral and radical cationic organic molecules
and of radical cations of push--pull azoheteroarene-based photoswitches.
These results importantly highlight the possibility
of achieving fast and efficient predictions of chemical properties
sensitive to the electronic structure
(\eg, charge carrier organic materials or transition-metal-catalyzed reaction steps).

\section{Theory}
\label{sec:theory}

This section provides a concise description of the proposed
atom-based \SPAHM[a] and bond-based \SPAHM[b] models
introduced in this work.
The general workflow used to generate the representation is sketched on Fig.~\ref{fig:theory},
and detailed derivations are shown in Sec.~\extref{sec:derivation} of the Supplementary Material.

\begin{figure*}
\centering
\includegraphics[width=\linewidth]{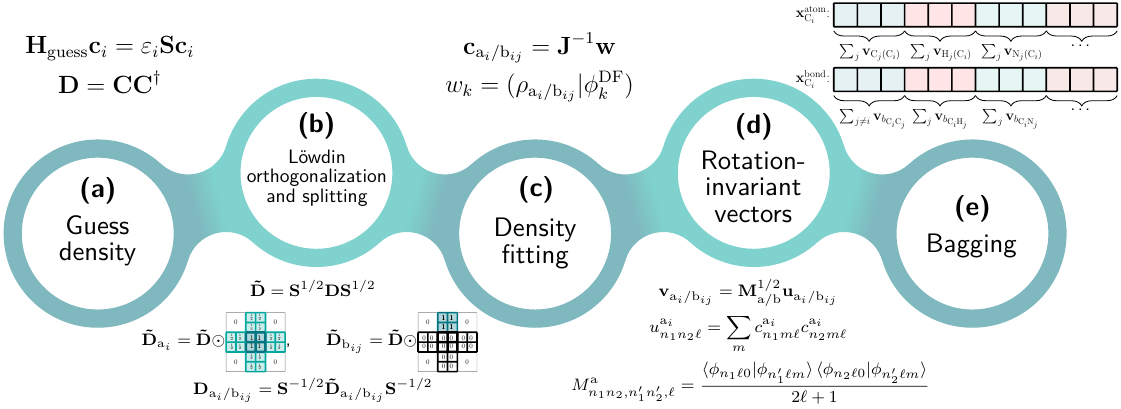}
\caption{
Scheme illustrating the steps required to compute \SPAHM[a] and \SPAHM[b] representations.
}
\label{fig:theory}
\end{figure*}

\subsection{\texorpdfstring{\SPAHM[a]}{SPAHM(a)}}
\label{sec:theory:a}

This work extends our former representation built from the eigenvalues of lightweight model Hamiltonians.
To achieve locality and transferability, the new representations focus on the eigenvectors of those Hamiltonians.
However, to avoid dealing with permutational invariance,
instead of the eigenvectors
our extension is based on the electron density $\rho(\vec r)$
or more specifically the pre-processed density matrix $\vec D$
(Fig.~\ref{fig:theory}a).

\smallskip

Local representations are designed to encode information
about each atom within a molecule into a vector.
It is thus natural to build our representation from the atomic electron density $\rho_I(\vec r)$ of each atom~$I$.
Yet, there is no unique way to attribute density to an atom.\cite{M1955,PM1971,L1950,H1977,BAAC2007,S1981,B1991,KP2000,NP2000,PBF2006}
Given the need for analytical solutions,
we choose to define $\rho_I(\vec r)$ in the form of a decomposition onto an atom-centered basis set
in the spirit of the density-fitting approximation.\cite{BER1973,W1973,ETOHA1995}

After performing L\"owdin orthogonalization\cite{L1950} of atomic orbitals,
we obtain a separate density matrix attributed to each atom (Fig.~\ref{fig:theory}b, left).
We then proceed with the density fitting
and can take the coefficients $\vec c_{I(I)}$
of the functions centered on the atom of interest (Fig.~\ref{fig:theory}c).

The density fitting step allows to
take into account the contribution of atoms ${J\neq I}$ to the $\rho_{I}(\vec r)$,
using the coefficients $\vec c_{J(I)}$
of decomposition of $\rho_I(\vec r)$ centered on nucleus $J$,
thus implicitly including bonding information.
The detailed description and comparison with other atomic partitioning schemes is provided in Sec.~\extref{sec:models}.
Note that in order to include bonding information explicitly,
the proposed approach can be generalized to obtain density matrices attributed to each bond (Sec.~\ref{sec:theory:b}).

\smallskip

The vector of coefficients $\vec c_I$ is not rotationally-invariant,
and hence cannot be directly exploited as a representation.
The next step corresponds to construction of a symmetry-adapted vector $\vec v_I$ (Fig.~\ref{fig:theory}d).

Inspired by the SOAP kernel,\cite{BKC2013}
we compute the similarity between two atoms
$A$ and $B$ as overlap of
$\rho_A(\vec r)$ and $\rho_B(\vec r)$.
To ensure rotational invariance, the overlap is integrated over all possible rotations in 3D space,
\begin{equation}
\kovlp{A,B} = \int \left| \braket{ \rho_A | \hat R | \rho_B} \right|^2 \de \hat R.
\end{equation}
(To obtain the overlap, the atoms $A$ and $B$ are virtually put at the same point of space.)
Note that this expression can be generalized
to ensure rotational \emph{equivariance} to learn
higher-order tensorial properties in spirit of $\lambda$-SOAP.\cite{GWCC2018}

Each atomic density $\rho_{I}(\vec r)$ is expressed as a sum of terms, centered on the nucleus $I$,
hence the overlap kernel can be written as a scalar product of two vectors,
$\kovlp{A,B} = \vec v_A\transp \vec v_B\phantomtransp$
(see Sec.~\extref{sec:derivation-a} for a detailed derivation of the expression for $\vec v_I$).
We disregard the overlap kernel
and use $\vec v_I$ as a representation vector of an atom.
It provides the following advantages:
\begin{enumerate*}[label=\roman*),itemjoin={{; }}]
\item
kernel computation is significantly simplified
\item
an atomic-density representation can be combined with other vectors
\item
the representations can be used with
any other kernel function, such as widely-used Laplacian and Gaussian kernels.
\end{enumerate*}

Another way to obtain a symmetry-adapted vector from $\vec c_I$,
reported in the context of ML density functionals,\cite{MR2021}
is to use sum of squares of density-fitting coefficients for each shell.
Comparison with our representation is provided in Sec.~\extref{sec:mr2021}.

\smallskip

The last step is to construct an atomic representation $\vec x_I$
from the symmetry-adapted vectors $\vec v_{J(I)}$.
We regroup all the vectors according to the charge of the nucleus $J$ into ``bags'' of element types,
inspired by the construction of aSLATM.\cite{HL2020}
Finally, we sum up the features in each bag to form the final vector.
This procedure is illustrated on Fig.~\ref{fig:theory}e (top).

\subsection{\texorpdfstring{\SPAHM[b]}{SPAHM(b)}}
\label{sec:theory:b}

As discussed in the previous section,
the bonding information is included only implicitly into \SPAHM[a].
A complementary approach consists of building an explicit representation
for a bond $IJ$ by extracting the corresponding density matrix and
the density $\rho_{IJ}(\vec r)$
with the L\"owdin formalism\cite{L1950}
(Fig.~\ref{fig:theory}b, right).

\smallskip

Using the standard density fitting approach,
$\rho_{IJ}$ could be expressed as a sum of terms centered on $I$ and $J$,
but this would preclude rewriting the kernel as a scalar product and then extracting a representation vector.
For this reason, we instead decompose $\rho_{IJ}(\vec r)$
onto a basis set centered in the middle of the $IJ$ bond
(Fig.~\ref{fig:theory}c).

Even though most of the information on the bond-density close to nuclei is lost during this procedure,
the behavior in the midbond region is well captured.
Bond-centered bases are
often used to extend atomic-orbital bases
for obtaining accurate interatomic potentials,\cite{TP1992,SH2018} but not for density fitting.
We thus optimized the basis for each bond present in the datasets studied
(involving elements H, C, N, O, F, S, see Sec.~\ref{sec:methods}).
The basis set construction is described in Sec.~\extref{sec:basis}.

\smallskip

Comparison of two bonds $AB$ and $CD$ involves
aligning them along the $z$-axis and superimposing
their geometrical centers.
The similarity is then computed
as an overlap of
$\rho_{AB}(\vec r)$ and $\rho_{CD}(\vec r)$,
integrated over the rotation around the $z$-axis (Fig.~\ref{fig:theory}d),
\begin{equation}
\kovlp{AB,CD}
= \int \left| \braket{ \rho_{AB} | \hat R_z | \rho_{CD}} \right|^2 \de \hat R_z
= \vec v_{AB}\transp \vec v_{CD}\phantomtransp.
\end{equation}
(See Sec.~\extref{sec:derivation-b} for a detailed derivation.)

Simplifications to reduce both the time
needed to compute the vector $\vec v_{IJ}$ and its size are possible.
For the fitting one can, for instance, use only basis functions
with magnetic quantum number $m=0$
to drop the integration over rotation around the $z$-axis,
or even leave only a single $s$- or $p$-orbital.
Sec.~\extref{sec:simplebond} illustrates how these simplifications
provide a useful compromise for certain datasets.

\smallskip

With the bond-representation vectors $\{\vec v_{IJ}\}$ at hand,
the similarity can be computed between two bonds.
While this could be readily used to train bond-property models
(\eg, bond dipole moments, dissociation energies)
this work focuses on atomic properties requiring one additional step to use the bond vectors and construct an atomic representation.

As for the atom-density representation (Sec.~\ref{sec:theory:a}),
we chose an aSLATM\cite{HL2020}-inspired ``bagging'' procedure (Fig.~\ref{fig:theory}e, bottom).
For each atom $A_i$, all the vectors $\vec v_{A_iB_j}$ are grouped
according to the element $B$ and summed up prior to concatenation.
Here, the difference between the bagging of \SPAHM[a] and \SPAHM[b]
is that the former is sorted according to unique elements (one-body terms in the language of SLATM)
and the latter --- according to pairs of unique elements (two-body terms).
This difference illustrates the complementary focus of the two new variations
of \SPAHM to convert the information from lightweight Hamiltonians into local atomic and bond environments.

\begin{figure}[hbt!]
\centering
\includegraphics[width=0.5\textwidth]{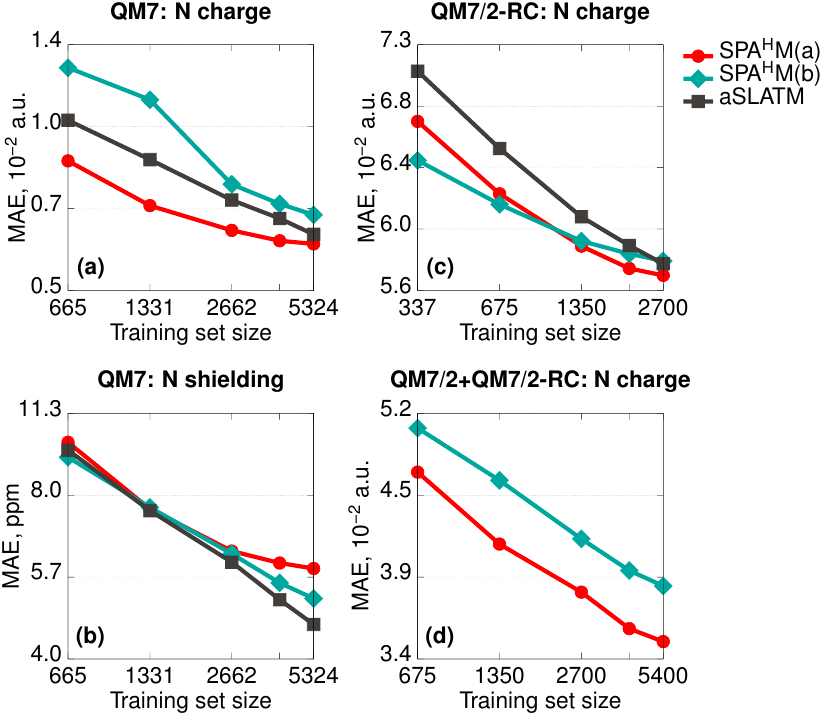}
\caption{
Learning curves for different datasets on the exemplary task of predicting local properties of nitrogen atoms:
\emph{(a)}~atomic charges and
\emph{(b)}~isotropic magnetic shielding constants for QM7
and atomic charges for
\emph{(c)}~radical cations of 3600 QM7 molecules (QM7/2-RC) and
\emph{(d)}~mix of 3600~QM7~molecules and 3600~radical cations (QM7/2+QM7/2-RC).
The QM7/2+QM7/2-RC aSLATM curve is missing since aSLATM is not injective and therefore inappropriate for this dataset.
}
\label{fig:QM7}
\end{figure}

\section{Results and discussion}

\subsection{Classic benchmark dataset: QM7}

We assess the learning ability of \SPAHM[a,b]
by predicting two distinct local atomic properties --- atomic charges and isotropic magnetic shielding constants ---
computed for the QM7 database\cite{RTML2012}.
For each element (H, C, N, O, S) and property, a separate kernel ridge regression (KRR)
model is trained using its own hyperparameters (see Sec.~\ref{sec:methods}).
Each set was randomly divided into a training and test set (\mbox{80\%\textrm{--}20\%}~split).

For each molecule, the LB\cite{LB2020} guess Hamiltonian paired with a minimal basis set\cite{K2013}
is diagonalized to obtain the atomic \SPAHM[a,b] representations
following the procedure described in Sec.~\ref{sec:theory} and Fig.~\ref{fig:theory}.
The LB guess was chosen owing to its best performance
for the eigenvalue-based \SPAHM{} (\eSPAHM)\cite{FBC2022}.
Comparison with the H\"uckel\cite{H1963,L2019} and PBE0\cite{AB1999} Hamiltonians are provided in Sec.~\extref{sec:potentials}.
Briefly, there is a correlation between the quality of the initial guess
and the performance of the representation,
which opens the way to improving \SPAHM[a,b]
through modifying the underlying guess Hamiltonian.

The learning curves of \SPAHM[a] and \SPAHM[b] for nitrogen atomic charges
are shown in Fig.~\ref{fig:QM7}a with comparisons with those of aSLATM\cite{HL2020}
(learning curves for other elements and properties are reported in Sec.~\extref{sec:lc-qm7}). \SPAHM[a] errors are comparable with those of aSLATM with no clear systematic trend across all the distinct elements
(see Sec.~\extref{sec:lc-qm7}).
The generally good performance of \SPAHM[a] arises from its well-suited atomic-density fingerprints,
which encode similar information to atomic charges.
Interestingly, the somewhat more sophisticated bond-variant \SPAHM[b] performs worse than \SPAHM[a], implying that the bonding information is less relevant for this task.
This contrasts with the predictions of isotropic shielding constants (Fig.~\ref{fig:QM7}b)
for which \SPAHM[b] is systematically superior to \SPAHM[a] owing to its dependence on the presence of multiple bonds
and $\pi$-conjugation, which are better captured by the bond density-based model.
Yet, for this property neither \SPAHM[a,b] outperform aSLATM.
Specifically, for the hydrogen atom (Sec.~\extref{sec:lc-qm7}),
most frequently analyzed in NMR studies of organic compounds,
the \SPAHM[b] error is $\sim1.5$ times higher than the aSLATM one.

This result is however not surprising as it was previously demonstrated with \eSPAHM
that the strength of the approach lies in capturing the properties of datasets
covering a broad range of chemical compositions and electronic structures featuring a variety of charges and spins.\cite{FBC2022}.
We thus train the model on two databases containing more electronically-diverse species.
The first one is made of radical cations of 3600~structures randomly selected from QM7 (QM7/2-RC),
and the second is a mixture of these 3600~neutral molecules and 3600~radical cations (QM7/2+QM7/2-RC).
The learning curves for nitrogen atomic charge
are shown on Fig.~\ref{fig:QM7}c and Fig.~\ref{fig:QM7}d, respectively.
For the QM7/2-RC dataset, we also predict atomic spin densities
(the complete set of learning curves available in Sec.~\extref{sec:lc-qm7}).
The \SPAHM[a,b] vectors for open-shell molecules are built from concatenation of vectors
obtained for $\alpha$ and $\beta$ densities (see Sec.~\extref{sec:alphabeta} for more details).

Fig.~\ref{fig:QM7}c illustrates the improved performance of \SPAHM[a,b] with respect to aSLATM,
for the difficult task of learning atomic charges of charged species.
Thanks to its rooting in the electron density, \SPAHM
is able to capture local changes in the electronic structure.
For QM7/2+QM7/2-RC (Fig.~\ref{fig:QM7}d),
aSLATM cannot be used as it yields the same representation vector
for a neutral molecule and its radical cation.
On the other hand, \SPAHM[a,b] seamlessly include the electronic information.
The overall prediction errors are approximately averaged errors for the QM7 and QM7/2-RC datasets.
However, \SPAHM[b] is always worse than \SPAHM[a] for both charges and spins for the same reason as discussed above.

\subsection{Tunable push--pull azoheteroarene-based dyes}

To assess the performance of \SPAHM[a,b] beyond prototypical molecular examples,
we consider a combinatorial database of push--pull azoheteroarene-based photoswitches\cite{VFBC2021} (APS),
containing 3429~molecules.
While this database was originally designed to analyze the tunability of excited states for this class of dyes,
we first investigate their hole-carrier properties and train predictive models
for the atomic charges and spins of radical cations (APS-RC).

\subsubsection{Predicting hole-carrier properties}

\begin{figure}[bh]
\centering
\includegraphics[width=0.5\textwidth]{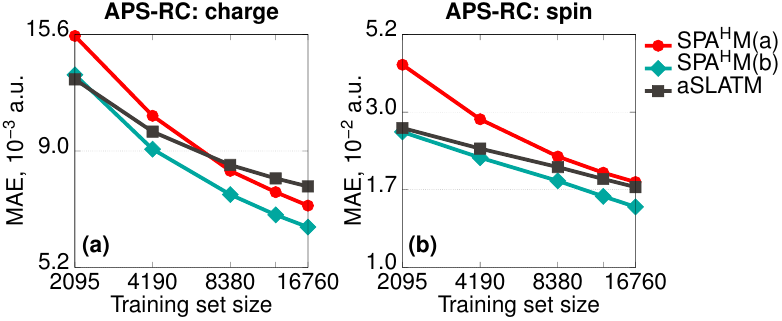}
\caption{
Learning curves of atomic charges and spins of~nitrogen
for the APS-RC dataset.
The reference properties are computed at the $\omega$B97X-D/def2-SVP level (see Sec.~\ref{sec:methods}).
}
\label{fig:azoswitch}
\end{figure}

For the APS-RC dataset,
the learning curves of \SPAHM[a,b] and aSLATM for nitrogen are shown in Fig.~\ref{fig:azoswitch}
(see Fig.~\extref{fig:lc-aps-rc} of Sec.~\extref{sec:lc-azoswitch} for other elements).
Akin to QM7/2-RC, \SPAHM[a] performs systematically better than aSLATM but \SPAHM[b] leads to the lowest errors for this set.
The superiority of \SPAHM[b] can be understood by taking a closer look at the chemical composition of the two sets.
QM7 consists of organic molecules with seven or less heavy atoms.
While it contains a large amount of structures with multiple and/or $\pi$-conjugated bonds,
there are only a few aromatic molecules and thus a restricted number
of fragments promoting extensive $\pi$-electron delocalization.
In contrast, the APS dyes are built from 2--4 donor--acceptor aromatic groups interacting through the azo moieties,
forming a fairly long $\pi$-conjugated scaffold prone to high charge and spin delocalization.
The bond-centered representation, which relies upon basis functions with components spatially orthogonal to the bond,
is suited to capture these electronic changes.
A deeper analysis of an individual molecule is provided in Sec.~\extref{sec:oos-aps-rc}.

\subsubsection{Predicting excited-state properties}

\begin{figure}[t]
\centering
\includegraphics[width=0.95\textwidth]{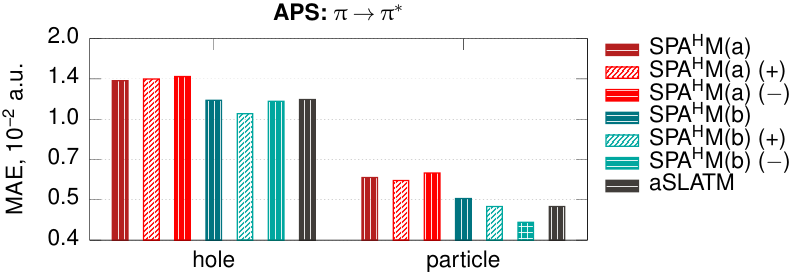}
\caption{
Prediction errors at the full training set
for contributions of nitrogen atoms to the hole and particle densities
of the productive $\pi\to\pi^*$ state for the APS dataset;
($+$) and ($-$) indicate \SPAHM computed for radical cations and anions, respectively.
The reference properties are computed with TDDFT at the $\omega$B97X-D/def2-SVP level (see Sec.~\ref{sec:methods}).
}
\label{fig:azoswitch-ex}
\end{figure}

Next, we challenged the representations with the APS dataset, considering a productive $\pi\to\pi^*$ excited state.
In line with the original work,\cite{VFBC2021} we focused on learning
atomic contributions to the hole and particle densities (computed in the same way as Hirshfeld charges).

\SPAHM is computed from a ground-state initial guess,
thus it cannot be expected to predict excited-state properties well.
Since the targets are atomic hole and particle contributions,
a reasonable approach is to use radical cation and anion densities, respectively, as a starting point.
Another choice would be to compute SPAHM from the guess HOMO and LUMO densities,
but it is not assessed here.
The prediction errors for neutral, cation, and anion \SPAHM[a,b]
and aSLATM for nitrogen are shown in Fig.~\ref{fig:azoswitch-ex}
(see Fig.~\extref{fig:lc-aps-ex} and Fig.~\extref{fig:lc-aps-ex-hist} of Sec.~\extref{sec:lc-azoswitch} for other elements).

Among the three \SPAHM density sources,
the anion one is systematically the best for the particle contributions,
and the cation one --- for the hole contributions,
while, as expected, the neutral one is usually the worst.

Only the anion-\SPAHM contains information on the LUMO,
which explains its good performance for the particle contributions
(particle density consists of unoccupied orbitals)
and low performance for the hole contributions
(hole density consists of occupied orbitals thus the LUMO information just adds extra noise).

The better performance of cation-\SPAHM in the case of hole contributions could be explained differently.
For open-shell systems, \SPAHM[a,b] consist of two concatenated vectors constructed from the $\alpha$- and $\beta$-densities.
Thus the full vector implicitly contains the information on the HOMO orbital density which
is removed from the $\beta$-vector with respect to $\alpha$.

In total, for the productive $\pi\to\pi^*$ state of the azo-photoswitches
we found the cation-\SPAHM[b] and anion-\SPAHM[b] to be the best for the hole and particle properties,
respectively, and expect the same trend for similar excited states.
For all the elements in the dataset this approach outperforms aSLATM,
which proves that the \SPAHM family can be useful also for excited-state properties.

\begin{figure}[bth]
\centering
\includegraphics[width=0.5\textwidth]{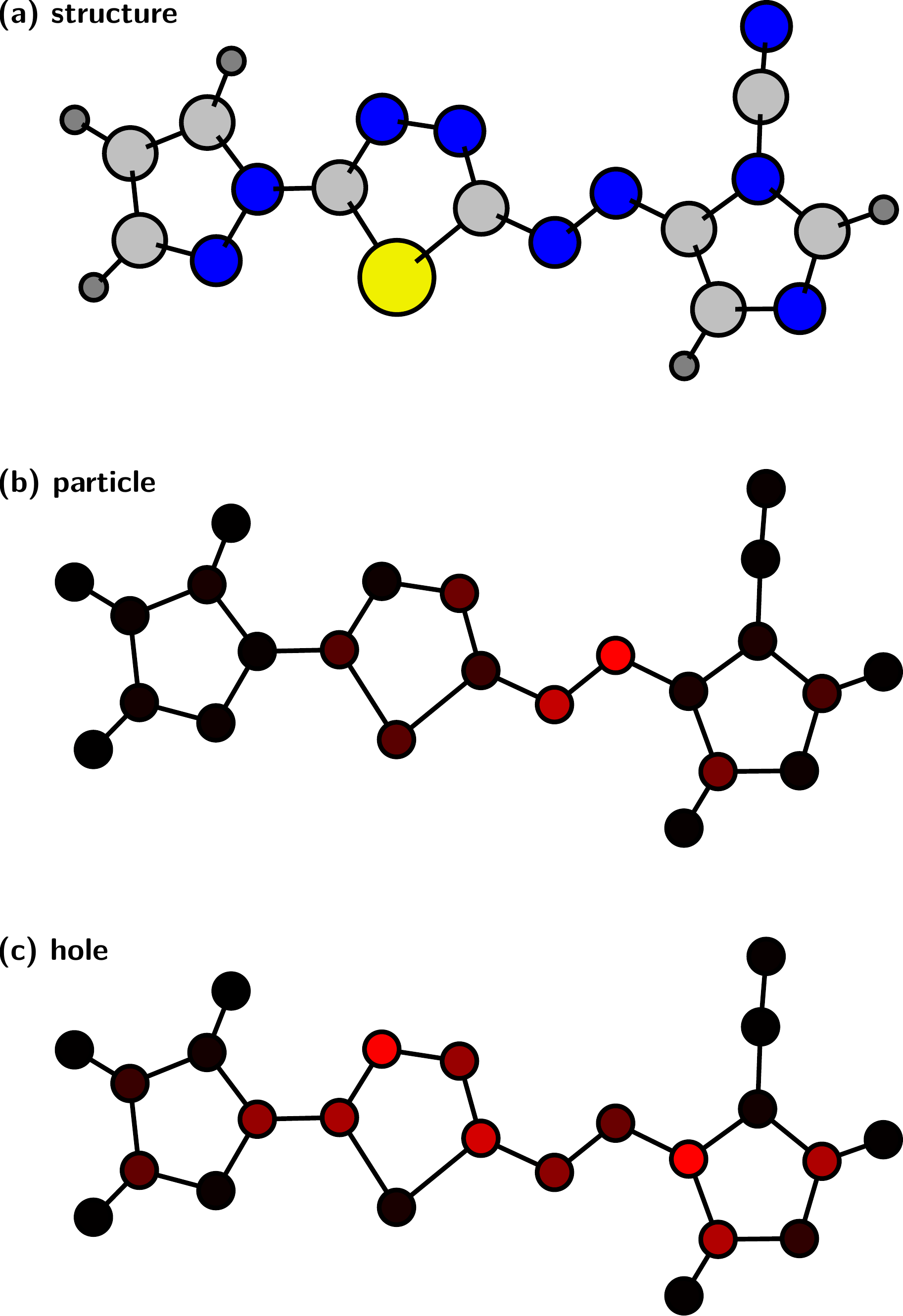}
\caption{
Qualitative picture
for \emph{(a)} an out-of-sample structure:
atomic contributions
to the \emph{(b)} particle and \emph{(c)} hole densities of the productive $\pi\to\pi^*$ state,
predicted by \SPAHM[b].
\emph{(a)}: The elements are color-coded by
dark gray for \ce{H}, light gray for \ce{C}, blue for \ce{N}, and yellow for \ce{S}.
\emph{(b,c)}: The contribution of each atom is represented by the
color intensity from black ($0$) to red (max.\ value).
}
\label{fig:oos-aps-ex}
\end{figure}

\subsubsection{Out-of-sample prediction}

We also predicted hole and particle contributions for an out-of-sample molecule,
for which a graphical representation is shown in Fig.~\ref{fig:oos-aps-ex},
and the numerical values are provided in Sec.~\extref{sec:oos-aps-ex}.
This structure is one of several that were excluded from training in the original work,\cite{VFBC2021}
because the computations yielded two almost degenerate $\pi\to\pi^*$ states of mixed character
which made it impossible to identify the target state. On the other hand, this makes it a good out-of-sample example,
because, in contrast to a quantum-chemical computation, the ML model does not know about other excited states
and thus can predict the properties of the ``correct'' state as if it existed.

The predicted values show a picture typical for this excitation:\cite{VC2020,VFBC2021}
the particle density is mostly localized on the azo group
(which makes it much easier to learn than the hole density).
Conversely, the hole density is delocalized all over the $\pi$-system,
its asymmetry showing the push--pull character of the excitation.


\begin{table*}[t]
\caption{\label{tab:time}User times required to generate the the \SPAHM[a], \SPAHM[b], and aSLATM
representations and to compute the Laplacian kernel
for the sub-QM7 and sub-APS sets (1000 randomly selected molecules). The values are averaged over 5~runs.
}

\begin{tabular}{@{}c c cccc c cccc c @{}}
\toprule
\multirow{3}{*}{Method}  & \multicolumn{5}{c}{sub-QM7}                                                        & \multicolumn{5}{c}{sub-APS}                                                         &  \multirow{3}{*}{\makecell{Repr.\\ size,\\ features}}  \\ \cmidrule(lr){2-6}\cmidrule(lr){7-11}
                         & \multirow{2}{*}{\makecell{Repr.,\\ h}} & \multicolumn{4}{c}{ Kernel, s}            & \multirow{2}{*}{\makecell{Repr.,\\ h}} &  \multicolumn{4}{c}{ Kernel, s}            &                                                        \\ \cmidrule(lr){3-6}\cmidrule(lr){8-11}
                         &                                        & \ce{H}   & \ce{C}   & \ce{N}   & \ce{O}   &                                        &  \ce{H}   & \ce{C}   & \ce{N}   & \ce{S}   &                                                        \\ \midrule
\SPAHM[a]                & 0.9                                    & \bf 1.37 & \bf 0.75 & \bf 0.07 & \bf 0.06 & 4.5                                    &  \bf 1.86 & \bf 6.08 & \bf 1.60 & \bf 0.09 & \bf 943                                                \\
\SPAHM[b]                & 8.2                                    & 3.36     & 1.25     & 0.14     & 0.06     & 42.0                                   &  3.54     & 5.79     & 1.73     & 0.05     & 1328                                                   \\
aSLATM                   & \bf 0.1                                & 27.9     & 10.1     & 0.34     & 0.31     & \bf 0.2                                &  53.9     & 93.3     & 26.1     & 0.75     & 10808                                                  \\
\bottomrule
\end{tabular}
\end{table*}

\subsection{Efficiency}
To complete this work, we evaluate the efficiency of our models compared to aSLATM,
for both the feature vectors generation and kernel computation.
Since the training of KRR models consists of the kernel matrix inversion,
which does not explicitly depend on the representation type,
the inversion times are not included.

We randomly selected a subset of 1000~molecules from the QM7 (sub-QM7)
and APS (sub-APS, fluorine-containing molecules excluded) databases
and generated the \SPAHM[a], \SPAHM[b], and aSLATM representations.
The user times are reported in Table~\ref{tab:time}.

\SPAHM[a,b] computation requires
diagonalization (per molecule) and density-fitting (per atom/bond in molecule) procedures,
in the worst case scaling cubically with the number of atoms,
resulting in being computationally expensive.
Compared to aSLATM, generating \SPAHM[a] vectors is approximately
9-fold more time-consuming (squared for \SPAHM[b] relatively) for the sub-QM7 dataset.
Moving toward more complex systems, \ie the sub-APS dataset,
reveals a larger observable time complexity than aSLATM: it took about twice more time to compute the aSLATM representation,
compared to QM7, and about five times for either \SPAHM[a,b].
However, the overall speed is implementation-dependent, and efficiency is being addressed in ongoing efforts.
Moreover, the simplified bond models discussed in Sec.~\extref{sec:simplebond}
open the route for future optimizations of \SPAHM[b].

We extend the analysis by computing the full Laplacian kernel matrices
from these representations (for each element separately),
the user times are reported in Table~\ref{tab:time}.
For a fixed set, the theoretical complexity of kernel computation is proportional
to the representation vector length (number of features).
Since the \SPAHM[a,b] vectors are $\sim 10$ times more compact than the aSLATM ones,
kernel computation for the former is significantly faster,
which is especially important for multiple runs needed for hyperparameter search.
While extension of \SPAHM[a,b] to open-shell systems leads to a two times increase
of the vector length, it is still linear with respect to the number of elements in the dataset
in contrast to the cubic dependency of aSLATM.

Thus, despite the computation of the \SPAHM[a,b] vectors requiring more time than the aSLATM ones,
training \SPAHM[a,b]+KRR models is more efficient than aSLATM+KRR.
This can be advantageous for molecular dynamics in active learning setups.\cite{ZLWCE2019}
We also note that in all cases the prediction time is negligible with respect to TDDFT computations,
whilst \SPAHM[b] shows good results for excited-state properties.
Finally, comparison of the original \SPAHM[b] with simplified versions
shows little deterioration of the overall performance and offers promising routes
toward more efficient implementations (See Sec.~\extref{sec:simplebond}).
Additionally, limiting the extent of the bond-based environments
by optimizing the cutoff distance would preclude the computation of distant pairs
while maintaining relevant motifs.
While being currently under investigation, this effort is expected
to significantly reduce run times especially for large systems.

\section{Conclusions}
This work expands our
lightweight and efficient eigenvalue \SPAHM representation into a local electron density-based variant.
The adopted strategy extends the class of fingerprints derived from an approximated Hamiltonian
with two local density-matrix-based representations: \SPAHM[a] and \SPAHM[b],
accounting for atom and bond contributions.

Combining strategies inspired from state-of-the-art local representations (\ie,
SOAP, aSLATM) while simultaneously encoding electronic information, the \SPAHM variants show
excellent predictive power on local atomic properties (\eg, atomic charges,
atomic spin density, and isotropic magnetic shielding) of neutral and charged
species for both the prototypical QM7 and more challenging
(azoheteroarene-based dyes) sets.
\SPAHM[a,b] were shown to outperform aSLATM for predicting properties of
cationic species generated from the QM7 database as well as for those of highly
conjugated cationic systems.
Validation on the azoheteroarene-based dye database also demonstrated that
\SPAHM[b] is especially adapted to describe changes in electron delocalization
typically observed in extended $\pi$-conjugated systems.

We note that \SPAHM[a] and \SPAHM[b] encode the electronic information while
retaining compactness with feature vectors about 4- and 9-fold smaller than
aSLATM, respectively. In particular, the size of the representations does not
depend on the molecular sizes in the dataset (\ie, the system size) but rather on the number of unique
elements contained in it. Detailed analysis of the efficiency of the models reveals that this
constitutes a significant advantage for the kernel construction.

Overall, the proposed representations afford a transferable (local) and efficient alternative
for quantum machine learning in the prediction of various electronic-state properties.
We also expect the new \SPAHM variants to provide a powerful and chemically intuitive
framework for the prediction of properties of chemical reactions,
which require a bond-focus\cite{GFWC2022} as found in \SPAHM[b],
and for the description of molecular properties for which geometrical structures do not
inherently coincide with electronic structures (\eg, organic electronic materials).

\section{Methods}
\label{sec:methods}

The codes used in this paper are available
on a dedicated GitHub repository at \url{https://github.com/lcmd-epfl/SPAHM-RHO}
and on \texttt{Q-stack}, a broader package for custom quantum-chemical routines to promote quantum machine learning,
at \url{https://github.com/lcmd-epfl/Q-stack}.

The initial guess densities were obtained in a minimal basis (MINAO\cite{K2013})
using the LBm potential\cite{LB2020,FBC2022}.
(Comparison with the H\"uckel\cite{H1963,L2019} and PBE0\cite{AB1999} potentials is provided in Sec.~\extref{sec:potentials}.)
To construct the \SPAHM[a] representation, the cc-pVDZ/JKFIT\cite{W2002,S2015} atom-centered density fitting basis was used.
To construct the \SPAHM[b] representation, a bond-centered density fitting basis was optimized,
the procedure is described in Sec.~\extref{sec:basis}.
The \texttt{QML}\cite{CFHBTMV2017} and \texttt{TENSOAP} (\texttt{SOAPFAST})\cite{WGAFNBFC2021} packages
were used to construct the aSLATM\cite{HL2020} and SOAP\cite{BKC2013,GWCC2018} representations, respectively.
The KDFA\cite{MR2021} representation, also used for comparison,
has been re-implemented by us based on the LB guess.

In this work, three molecular datasets were used.
They were divided into atomic datasets for each element
and randomly split into training and test sets (80\%--20\%):
\begin{enumerate*}[label=\roman*),itemjoin={{; }}, itemjoin*={{ and }}]
\item QM7\cite{RTML2012} (7165~neutral organic molecules
containing \num{61959}~\ce{H}, \num{35761}~\ce{C}, \num{6655}~\ce{N}, \num{5978}~\ce{O}, and \num{297}~\ce{S})
\item QM7/2-RC (radical cations of 3600~randomly selected structures from QM7,
containing \num{31195}~\ce{H}, \num{17946}~\ce{C}, \num{3375}~\ce{N}, \num{3020}~\ce{O}, and \num{152}~\ce{S})
\item APS
\item \mbox{APS-RC}
(3429~azo-photoswitches,\cite{VFBC2021}
containing \num{29526}~\ce{H}, \num{39551}~\ce{C}, \num{20951}~\ce{N}, \num{1053}~\ce{O}, \num{741}~\ce{F}, and \num{3337}~\ce{S},
and the corresponding radical cations)
\end{enumerate*}.

The atomic charges and spins and/or hole and particle contributions
were computed using dominant Hirshfeld partitioning\cite{H1977}
at the PBE0\cite{AB1999}/cc-pVQZ\cite{D1989,WD1993} level
for the QM7\cite{RTML2012} and QM7/2-RC datasets
and at the $\omega$B97X-D\cite{CH2008}/def2-SVP\cite{WA2005} level
for the APS\cite{VFBC2021} and APS-RC datasets.
The excited-state properties were computed with TDDFT within the Tamm--Dancoff approximation.\cite{HH1999}
The isotropic shielding constants were computed
at the PBE\cite{PBE1996,PBE1997}/cc-pVDZ\cite{D1989} level.
All quantum-chemical computations were made with \texttt{PySCF 2.0}.\cite{S2015,SBBBGLLMSSWC2017}

For each dataset, element, property, and representation,
a separate kernel ridge regression (KRR)
model is trained using its own hyperparameters.
The hyperparameters (kernel type, kernel width, regularization)
were optimized with a grid search using a 5-fold cross-validation procedure
and the learning curves were computed with random sub-sampling (5~times per point).
The optimization and regression codes use
the \texttt{numpy}\cite{HMWGVCWTBSKPHKBHFWPGSRWAGO2020}
and \texttt{scikit-learn}\cite{PVGMTGBPWDVPCBPD2011} \texttt{python} libraries.
The optimal hyperparameters
can be found in the GitHub repository (\url{https://github.com/lcmd-epfl/SPAHM-RHO})
as well as in Materials Cloud (\url{https://doi.org/10.24435/materialscloud:1g-w5})
together with the learning curves.


\section*{Assosiated content}

\subsection*{Data availability}
The data and the model that support the findings of this study
are freely available in Materials Cloud (\url{https://doi.org/10.24435/materialscloud:1g-w5}).

The code is available in \texttt{Q-stack}
(\url{https://github.com/lcmd-epfl/Q-stack})
and as a separate GitHub repository at
\url{https://github.com/lcmd-epfl/SPAHM-RHO}.

\subsection*{Supplementary material}
Supplementary material contains:
\begin{enumerate*}[label=\roman*),itemjoin={{; }}]
\item
detailed derivation of \SPAHM[a,b] overlap kernels
\item
learning curves for all the elements
\item
predictions for the APS-RC and APS out-of-sample system
\item
comparisons of guess Hamiltonians,
different atom-density-based models, and
generalizations to open-shell systems
\item
\SPAHM[b] basis set choice discussion
\item
comparison with the KDFA\cite{MR2021} representation
\end{enumerate*}.

\section*{Author information}

\subsection*{Author contributions}
K.~R.~B. and Y.~C.~A. contributed equally to this work.
K.~R.~B. and Y.~C.~A. performed the computations and developed the software.
K.~R.~B., Y.~C.~A., A.~F., and C.~C. designed the representations and conceptualized the project.
All the authors contributed to the writing, reviewing, and editing of the manuscript.
C.~C. is credited for funding acquisition.

\subsection*{Notes}
The authors declare no competing financial interest.

\section*{Acknowledgements}
The authors acknowledge
the European Research Council (grant number~817977),
the NCCR MARVEL, a National Centre of Competence in Research,
funded by the Swiss National Science Foundation (grant number~205602),
and the EPFL for financial support.
The authors thank Puck van Gerwen and Rub\'{e}n Laplaza for helpful discussions.

\bibliography{spahm+.bib}
\clearpage

\begin{tocentry}
\centering{
\includegraphics[]{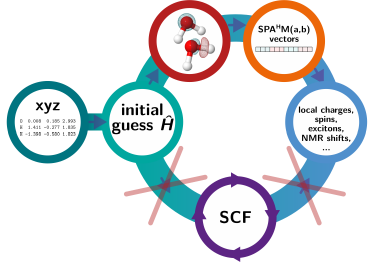}
}
\end{tocentry}

\end{document}


\title{{\sc Supplementary Information}\texorpdfstring{\\}{}
\texorpdfstring{\SPAHM[a,b]}{SPAHM(a,b)}:
Encoding the density information from guess Hamiltonian in quantum machine learning representations}

\author{Ksenia R. Briling}
\affiliation{Laboratory for Computational Molecular Design, Institute of Chemical Sciences and Engineering,
\'{E}cole Polytechnique F\'{e}d\'{e}rale de Lausanne, 1015 Lausanne, Switzerland}
\author{Yannick Calvino Alonso}
\affiliation{Laboratory for Computational Molecular Design, Institute of Chemical Sciences and Engineering,
\'{E}cole Polytechnique F\'{e}d\'{e}rale de Lausanne, 1015 Lausanne, Switzerland}
\author{Alberto Fabrizio}
\affiliation{Laboratory for Computational Molecular Design, Institute of Chemical Sciences and Engineering,
\'{E}cole Polytechnique F\'{e}d\'{e}rale de Lausanne, 1015 Lausanne, Switzerland}
\affiliation{National Centre for Computational Design and Discovery of Novel Materials (MARVEL),
\'{E}cole Polytechnique F\'{e}d\'{e}rale de Lausanne, 1015 Lausanne, Switzerland}
\author{Clemence Corminboeuf}
\email{clemence.corminboeuf@epfl.ch}
\affiliation{Laboratory for Computational Molecular Design, Institute of Chemical Sciences and Engineering,
\'{E}cole Polytechnique F\'{e}d\'{e}rale de Lausanne, 1015 Lausanne, Switzerland}
\affiliation{National Centre for Computational Design and Discovery of Novel Materials (MARVEL),
\'{E}cole Polytechnique F\'{e}d\'{e}rale de Lausanne, 1015 Lausanne, Switzerland}

\date{\today}

\maketitle
\onecolumngrid
\tableofcontents
\clearpage

\section{Derivation of the \texorpdfstring{\SPAHM[a,b]}{SPAHM(a,b)} overlap kernels}
\label{sec:derivation}
\subsection{Atom density [\texorpdfstring{\SPAHM[a]}{SPAHM(a)}]}
\label{sec:derivation-a}

Let us consider two atoms, $A$ and $B$.
Each atomic density $\rho_I(\vec r)$ is represented
as a linear combination of atom-centered spherical Gaussian basis functions $\{\phi_{n\ell m}\}$,
labeled with their radial channel number $n$ and angular $\ell$ and magnetic $m$ quantum numbers,
\begin{gather}
\rho_A(\vec r) = \sum_{n\ell m} c_{n\ell m}^A \phi_{n\ell m}(\vec r), \quad
\rho_B(\vec r) = \sum_{n'\ell'm'} c_{n'\ell'm'}^B \phi_{n'\ell'm'}(\vec r),
\end{gather}
and each nucleus is virtually positioned at the origin.

The overlap kernel $\kovlp{A,B}$ between atoms $A$ and $B$ is the squared overlap
of $\rho_A$ and $\rho_B$ averaged over all possible relative orientations $\hat R$,
\begin{equation}
\kovlp{A,B}
= \frac{1}{8\pi^2}\int \left|\braket{\rho_A|\hat R|\rho_B}\right|^2 \de \hat R
= \frac{1}{8\pi^2}\int \left|k_{AB}(\hat R)\right|^2 \de \hat R.
\end{equation}

For a given orientation, the overlap $k_{AB}(\hat R)$ is
\begin{equation}
\begin{split}
k_{AB}(\hat R)
&= \braket{\rho_A|\hat R|\rho_B}
=\int \rho_A(\vec r) \hat R\, \rho_B(\vec r) \de^3 \vec r
\\&=\sum_{n\ell m} c_{n\ell m}^A \sum_{n'\ell'm'} c_{n'\ell'm'}^B
\braket{\phi_{n\ell m}| \hat R\, \phi_{n'\ell'm'}}
\\&=\sum_{n\ell m} c_{n\ell m}^A \sum_{n'\ell'm'} c_{n'\ell'm'}^B
\Braket{\phi_{n\ell m}| \sum_{m''} \phi_{n'\ell'm''} D_{m''m'}^{\ell'}(\hat R)}
\\&=\sum_\ell\sum_{nm} \sum_{n'm'} c_{n\ell m}^A c_{n'\ell m'}^B
A_{nn'}^\ell D_{mm'}^{\ell}(\hat R),
\end{split}
\end{equation}
where $\vec D$ are Wigner D-matrices for \emph{real} spherical harmonics\cite{VMK1988}
and $A_{nn'}^\ell = \braket{\phi_{n\ell m} |\phi_{n'\ell m}} \ \forall m$.

The kernel becomes
\begin{equation}
\begin{split}
\kovlp{A,B}
&=\frac{1}{8\pi^2} \int \Big|
\sum_\ell\sum_{nm}\sum_{n'm'} c_{n\ell m}^A c_{n'\ell m'}^B
A_{nn'}^\ell D^{\ell}_{mm'}(\hat R)
\Big|^2 \de \hat R
\\&=
\frac{1}{8\pi^2}\sum_{\substack{{\ell_1\,n_1m_1\,n'_1m'_1}\\{\ell_2\,n_2m_2\,n'_2m'_2}}}
c_{n_1\ell_1m_1}^A c_{n'_1\ell_1m'_1}^B
c_{n_2\ell_2m_2}^A c_{n'_2\ell_2m'_2}^B
A_{n_1n'_1}^{\ell_1} A_{n_2n'_2}^{\ell_2}
\cdot\int
D^{\ell_1}_{m_1m'_1}(\hat R)\,
D^{\ell_2}_{m_2m'_2}(\hat R)\,\de \hat R.
\end{split}
\end{equation}
Thanks to orthogonality of the real Wigner D-matrices,\cite{VMK1988} \ie,
\begin{equation}
\int
D^{\ell_1}_{m_1m_1'}(\hat R)\,
D^{\ell_2}_{m_2m_2'}(\hat R)\,
\de \hat R
=
\frac{8\pi^2}{2\ell_1+1}\,\delta_{\ell_1\ell_2}\delta_{m_1m_2}\delta_{m'_1m'_2},
\end{equation}
the kernel is further simplified to
\begin{equation}
\label{eq:atom-kernel-full}
\kovlp{A,B} =
\sum_{\ell}  \sum_{\substack{n_1n'_1 \\ n_2n'_2 }}
\underbrace{
\left(\sum_m    c_{n_1\ell m}^A   c_{n_2\ell m}^A\right)
}_{u^A_p}
\underbrace{
\left(\frac{A_{n_1n'_1}^{\ell} A_{n_2n'_2}^{\ell}}{2\ell+1}\right)
}_{M_{pq}}
\underbrace{
\left(\sum_m c_{n'_1\ell m}^B c_{n'_2\ell m}^B\right)
}_{u^B_q}.
\end{equation}
With $p=(n_1,n_2,\ell), q=(n'_1,n'_2,\ell)$ it can be rewritten as a dot product
\begin{equation}
\label{eq:atom-kernel-brief}
\kovlp{A,B}
= \sum_{pq} u_p^A M_{pq} u_q^B = \vec u_A\transp \vec M \vec u_B \phantomtransp
= (\vec M^{1/2} \vec u_A)\transp (\vec M^{1/2} \vec u_B) = \vec v_A\transp \vec v_B\phantomtransp,
\end{equation}
where $\vec v_I$ is the representation of an atomic electron density $\rho_I(\vec r)$
and is an analog of the power spectrum of atomic neighbor density.\cite{BKC2013}

\clearpage
\subsection{Bond density [\texorpdfstring{\SPAHM[b]}{SPAHM(b)}]}
\label{sec:derivation-b}

Now let us consider two bonds, $AB$ and $XY$.
The (L\"owdin) bond densities $\rho_{AB}(\vec r)$ and $\rho_{XY}(\vec r)$ are
decomposed onto basis sets centered in the middle of each bond,
\begin{equation}
\label{eq:bond-fit}
\rho_{AB}(\vec r) = \sum_{i} c_{i} \phi_i(\vec r), \quad
\rho_{XY}(\vec r) = \sum_{j} c_{j} \phi_j(\vec r),
\end{equation}
where a function $\phi_i$ is defined by a radial channel number $n_i$ and angular $\ell_i$ and magnetic $m_i$ quantum numbers.
Both bonds are aligned along the $z$-axis and their midpoints are put at the origin.

The overlap kernel $\kovlp{AB,XY} = \mathcal I_1$ between the two bonds $AB$ and $XY$ is defined
as a overlap integral $\mathcal I_2(\varphi)$
squared averaged over rotations $\hat \varphi_z$ around the $z$-axis,
\begin{equation}
\mathcal I_1
= \frac1{2\pi}\int_0^{2\pi} \de \varphi \left| \braket{\rho_{AB}| \hat \varphi_z| \rho_{XY}} \right|^2
= \frac1{2\pi}\int_0^{2\pi} \de \varphi \left| \mathcal I_2(\varphi) \right|^2.
\end{equation}

With the decomposition~\eqref{eq:bond-fit}, the overlap integral $\mathcal I_2(\varphi)$
is rewritten with overlap of the basis functions,
\begin{equation}
\mathcal I_2(\varphi) = \braket{ \rho_{AB} | \hat \varphi_z | \rho_{XY} }
= \sum_{ij} c_{i} c_{j} \braket{\phi_i | \hat \varphi_z | \phi_j }
= \sum_{ij} c_{i} c_{j} \mathcal I_3^{ij}(\varphi),
\end{equation}
as well as the kernel $\mathcal I_1$,
\begin{equation}
\mathcal I_1
= \frac{1}{2\pi} \sum_{iji'j'} c_{i} c_{j} c_{i'} c_{j'}
\int \mathcal I_3^{ij}(\varphi)\mathcal I_3^{i'j'}(\varphi) \de \varphi
= \sum_{iji'j'} c_{i} c_{j} c_{i'} c_{j'} \mathcal I_6^{iji'j'}.
\end{equation}

With the rules for rotation of real spherical harmonics around the quantization axis,
the overlap $\mathcal I_3(\varphi)$ becomes
\begin{equation}
\mathcal I_3^{ij}(\varphi)
= \braket{\phi_i | \hat \varphi_z | \phi_j}
= \braket{\phi_i | \phi_{j}     }\cos {m_j\varphi}
+ \braket{\phi_i | \phi_{\bar \jmath}}\sin {m_j\varphi}
= S_{ij     }\cos {m_j\varphi} + S_{i\bar\jmath}\sin {m_j\varphi},
\end{equation}
where $\phi_{\bar \jmath}$ is the same basis function as $\phi_{j}$ but with an opposite phase
(\ie $m_j = -m_{\bar \jmath}$).
The integral over rotation $\mathcal I_6$ is simplified to
\begin{equation}
\mathcal I_6^{iji'j'} =
\delta_{|m_j|,|m_{j'}|} (S_{ij} S_{i'j'} + S_{i\bar\jmath} S_{i'\bar\jmath'}(1-\delta_{m_j0}) ),
\end{equation}
and the overlap kernel $\mathcal I_1$ --- to
\begin{equation}
\mathcal I_1
= \sum_{iji'j'} c_{i} c_{j} c_{i'} c_{j'}
\delta_{|m_j|,|m_{j'}|} (S_{ij} S_{i'j'} + S_{i\bar\jmath} S_{i'\bar\jmath'}(1-\delta_{m_j0}) ).
\end{equation}

When $p$ and $q$ are centered at the same point,
$S_{pq} = \delta_{\ell_p \ell_q} \delta_{m_p m_q} A_{n_p n_q}^{\ell_p}$.
Thus $\mathcal I_1$ is further simplified to
\begin{equation}
\label{eq:bond-kernel-full}
\mathcal I_1
= \sum_{ii'jj'}
\underbrace{(c_{i} c_{i'} \delta_{|m_i|,|m_{i'}|} )}_{u^{AB}_{ii'}}
\underbrace{
\delta_{\ell_{i}  \ell_{j}}    A_{n_{i}  n_{j}} ^{\ell_{i}}
\delta_{\ell_{i'} \ell_{j'}}   A_{n_{i'} n_{j'}}^{\ell_{i'}}
( \delta_{m_{i}  m_{j}}  \delta_{m_{i'} m_{j'}}
+  \delta_{m_{i} , -m_{j }} \delta_{m_{i'}, -m_{j'}} (1-\delta_{m_j0}) )}_{M_{ii',jj'}}
\underbrace{(c_{j} c_{j'} \delta_{|m_j|,|m_{j'}|} )}_{u^{XY}_{jj'}},
\end{equation}
which can be rewritten as a dot product in the same spirit as the atom-density kernel,
\begin{equation}
\label{eq:bond-kernel-brief}
\kovlp{AB,XY}
= \sum_{ii'jj'} u_{ii'}^{AB} M_{ii',jj'} u_{jj'}^{XY}
= \vec u_{AB}\transp \vec M \vec u_{XY} \phantomtransp
= (\vec M^{1/2} \vec u_{AB})\transp (\vec M^{1/2} \vec u_{XY})
= \vec v_{AB}\transp \vec v_{XY} \phantomtransp,
\end{equation}
where $\vec v_{IJ}$ is the representation of a bond density $\rho_{IJ}(\vec r)$.

\clearpage


\section{Learning curves}
\label{sec:lc}

\subsection{QM7 and its derivatives}
\label{sec:lc-qm7}
\begin{figure}[h]
\centering
\includegraphics[width=1.0\linewidth]{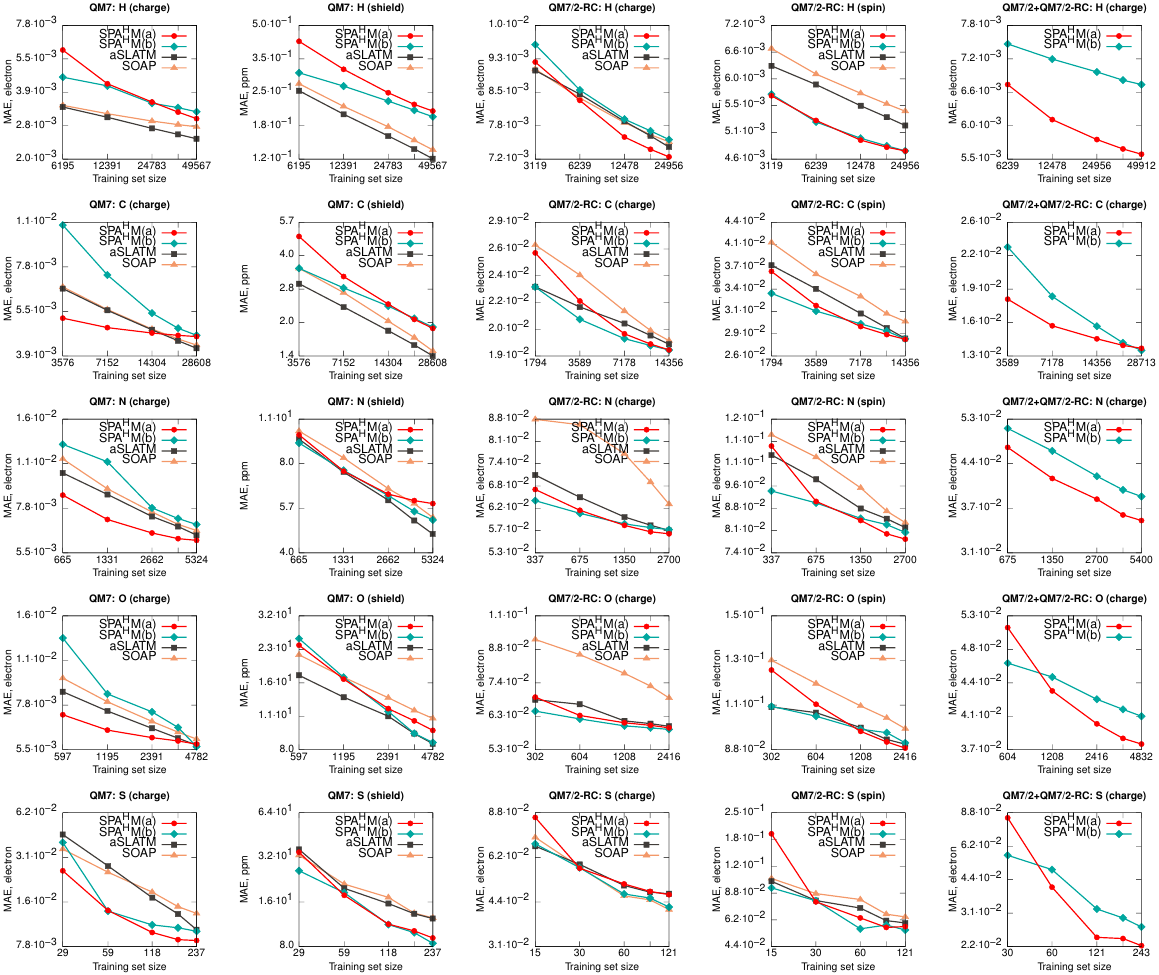}
\caption{
Learning curves of atomic charges and spins
for the QM7, QM7/2-RC, and QM7/2+QM7/2-RC datasets.
}
\label{fig:lc-qm7}
\end{figure}

\clearpage 

\subsection{APS-RC and APS}
\label{sec:lc-azoswitch}

\begin{figure}[h!]
\centering
\includegraphics[width=0.45\linewidth]{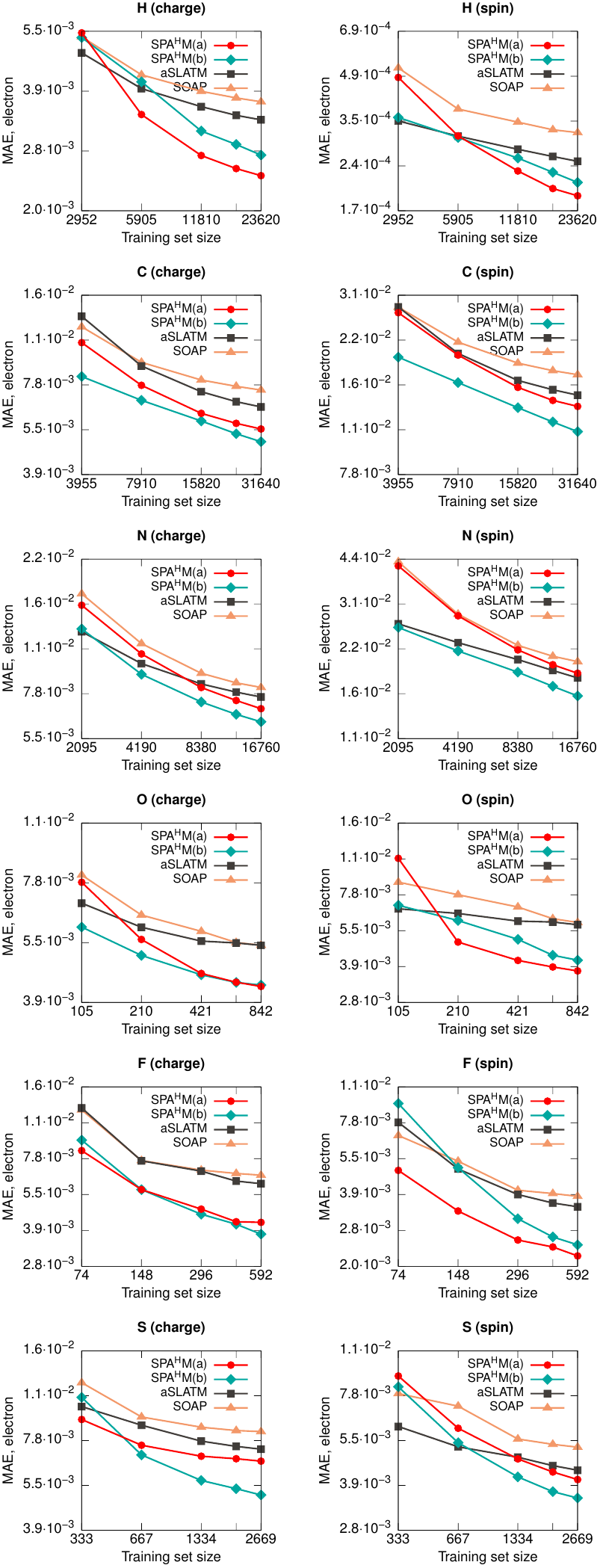}
\caption{
Learning curves of atomic charges and spins for the APS-RC dataset.
}
\label{fig:lc-aps-rc}
\end{figure}

\clearpage

\begin{figure}[ht]
\centering
\includegraphics[width=0.48\linewidth]{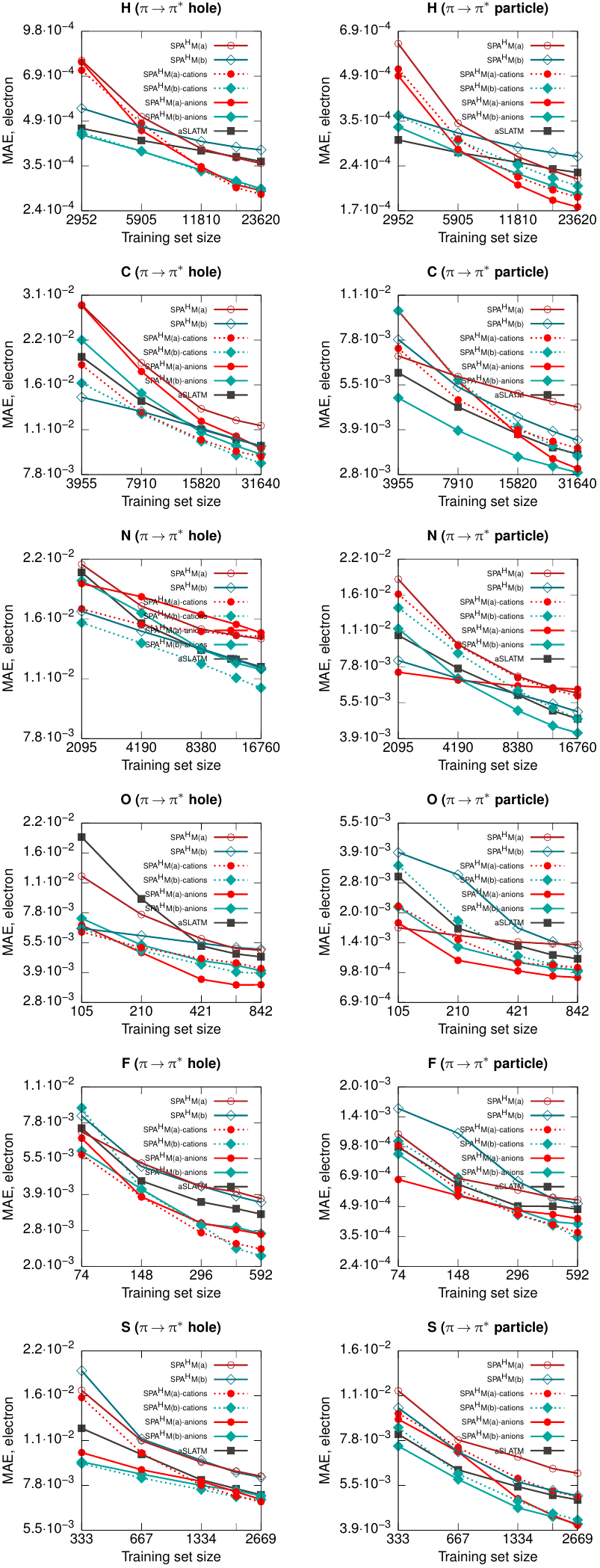}
\caption{
Learning curves of atomic contributions
to the hole and particle densities of the productive $\pi\to\pi^*$ state
for the APS dataset;
($+$) [dashed line] and ($-$) [solid line]
indicate \SPAHM computed for radical cations and anions, respectively.
}
\label{fig:lc-aps-ex}
\end{figure}

\clearpage

\begin{figure}[ht]
\centering
\includegraphics[width=0.38\linewidth]{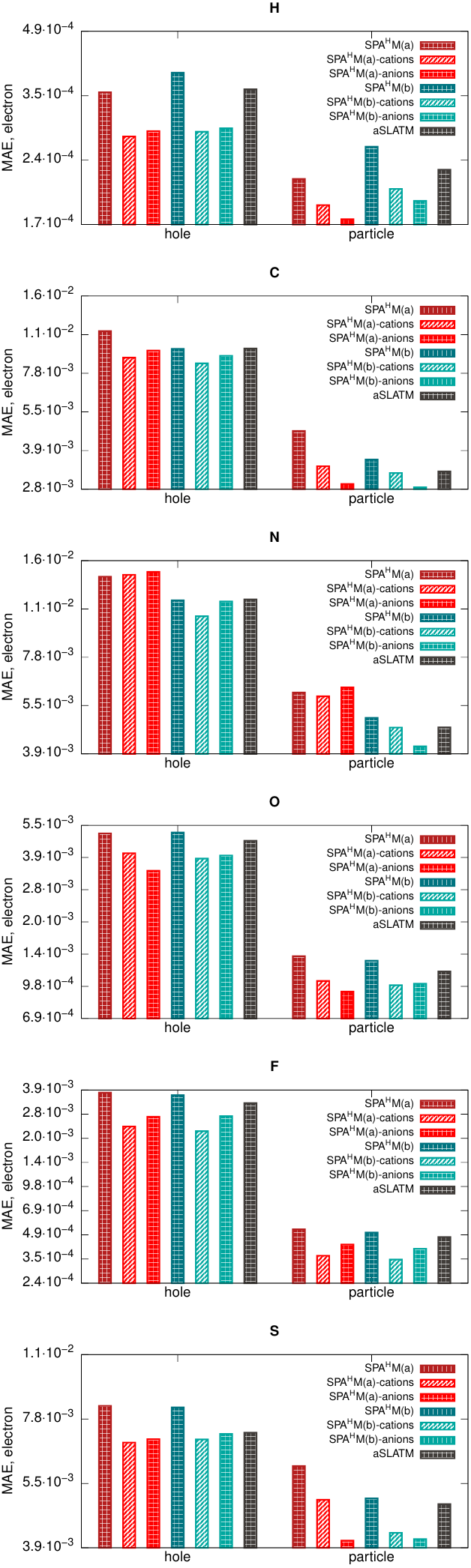}
\caption{
Histograms of the full training set errors
of atomic contributions
to the hole and particle densities of the productive $\pi\to\pi^*$ state
for the APS dataset;
($+$) [dashed line] and ($-$) [solid line]
indicate \SPAHM computed for radical cations and anions, respectively.
}
\label{fig:lc-aps-ex-hist}
\end{figure}

\clearpage


\section{Out-of-sample system}

\subsection{APS-RC}
\label{sec:oos-aps-rc}

\begin{figure}[ht]
\centering
\includegraphics[width=.5\linewidth]{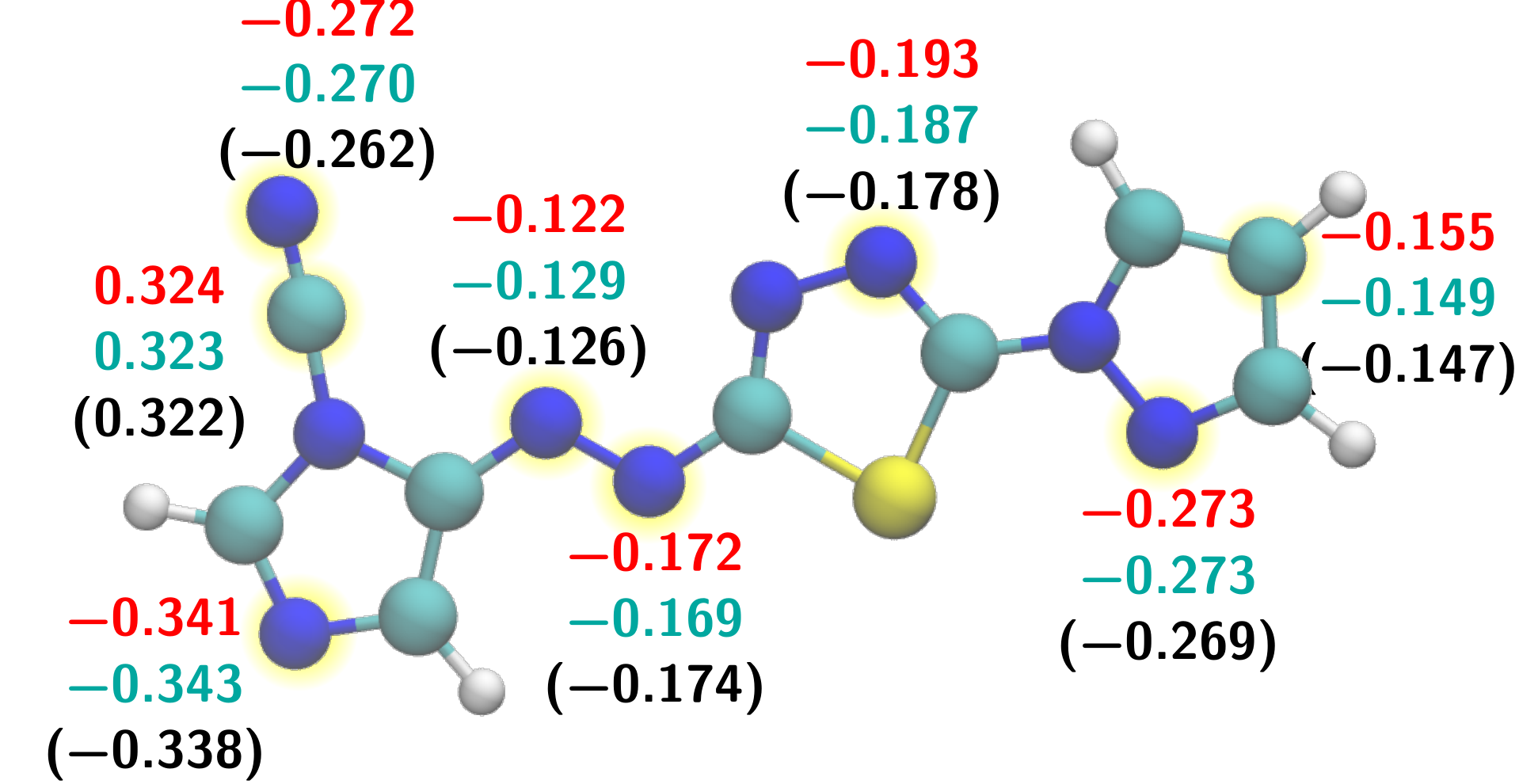}
\caption{
Predicted by \SPAHM[a] (red) and \SPAHM[b] (blue) and computed (black) atomic charges
for a radical cation of an out-of-sample structure on a selection of atoms (highlighted).
}
\label{fig:oos-aps-rc}
\end{figure}

Analysis of an individual system clearly illustrates the relevance of our models.
From the APS-RC dataset we selected an out-of-sample structure and
used previously trained \SPAHM[a,b] models to predict the atomic charges of its radical cation.
Fig.~\ref{fig:oos-aps-rc} compares the predicted and computed values
of atomic charges for a selection of atoms included in the $\pi$-conjugated system.
For \SPAHM[b], the predicted values accurately reproduce the computed ones within \SI{0.01}{{a.u.}},
thus verifying its performance.
However, by taking the changes in atomic charges for all the constituing atoms
and summing them up (\ie $\sum_{k}(q_k^\mathrm{cation}-q_k^\mathrm{neutral})$)
we obtain a total molecular charge $\sim 0.9$,
approximately yielding the removed electron.

\subsection{APS}
\label{sec:oos-aps-ex}
\begin{figure}[ht]
\centering
\includegraphics[width=.85\linewidth]{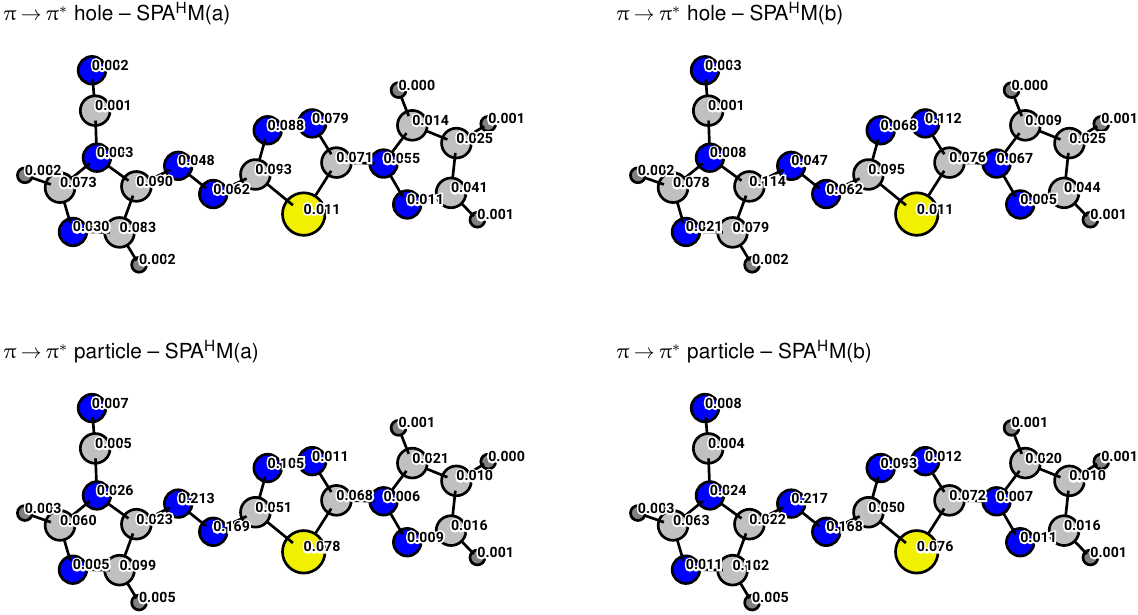}
\caption{
Predicted by \SPAHM[a,b] atomic contributions
to the hole and particle densities of the productive $\pi\to\pi^*$ state
for an out-of-sample structure.
}
\label{fig:oos-aps-ex}
\end{figure}

\clearpage


\section{Comparison of different atom-density-based models}
\label{sec:models}

In this section we describe and compare four models
used to post-process the guess density matrix.
The key elements of all of them are
density fitting\cite{BER1973,W1973,ETOHA1995} (DF),
\ie decomposition of the electron density onto an atom-centered basis set,
\begin{equation}
\vec c = \vec J^{-1} \vec w,
\quad
w_i = \sum_{pq} D_{pq} (\chi_p \chi_ q| \phi_i),
\end{equation}
where
$\vec D$ is a density matrix,
$\{\chi_p\}$ is the atomic orbital basis,
$\{\phi_i\}$ is the density-fitting basis,
$J_{ij} = (\phi_i | \phi_j)$,
and $(\cdots|\cdots)$ is a two-electron integral in chemists' notation,
and a subsequent symmetrization described in Sec.~\ref{sec:derivation-a}.

\begin{itemize}
\item The \emph{pure} model simply consists of fitting the guess density
      and partitioning of the resulting vector according to the nuclei centers
      following by symmetrization,
      \begin{equation}
      \vec D^\mathrm{guess}
      \becomes{DF} \vec c
      \becomes{part.} \{\vec c_I\}
      \becomes{sym.} \{\vec v_I\},
      \end{equation}
      (\ie $c_i \in \vec c_I$ if $\phi_i$ is centered on the nuclei $I$).

\item The \emph{diff} model consists of the same steps except that
      the difference between the guess density and the superposition of atomic densities (SAD) is used,
      \begin{equation}
      \vec D^\mathrm{guess}-\vec D^\mathrm{SAD}
      \becomes{} \vec c
      \becomes{} \{\vec c_I\}
      \becomes{} \{\vec v_I\}.
      \end{equation}
\end{itemize}

Both the \emph{short} and \emph{long} models follow the L\"owdin population analysis\cite{L1950}
to partition the molecular density matrix into atomic contributions $\{\vec D_{(I)}\}$,
\begin{equation}
\vec D^\mathrm{guess}
\becomes{}  \tilde{\vec D} = \vec S^{1/2} \vec D \vec S^{1/2}
\becomes{}  \{\tilde{\vec D}_{(I)}\}
\becomes{}  \{\vec D_{(I)} = \vec S^{-1/2} \tilde{\vec D}_{(I)} \vec S^{-1/2}\},
\end{equation}
where $\vec S$ is the atomic orbitals overlap matrix.
The resulting atomic density matrices $\{\vec D_{(I)}\}$
are individually subject to density fitting and symmetrization.

\begin{itemize}
\item The \emph{long} version
      includes the coefficients related to other atom centers
      as a long-range contribution to the atomic density:
      \begin{equation}
      \vec D^\mathrm{guess}
      \becomes{L\"owdin} \vec D^\mathrm{guess}_{(I)}
      \becomes{DF} \vec c_{(I)}
      \becomes{part.} \{\vec c_{J(I)}\}
      \becomes{sym.} \{\vec v_{J(I)}\}
      \quad \forall\,I.
      \end{equation}
      To construct the final representation for atom $I$, the vectors $\{\vec v_{J(I)}\}$
      are grouped according to the nuclear charge of $J$, summed up, and concatenated,
      but it is not the only possible way to proceed.

\item The \emph{short} version only retains the coefficients directly related
      to the basis functions centered on the atom of interest,
      \begin{equation}
      \vec D^\mathrm{guess}
      \becomes{} \vec D^\mathrm{guess}_I
      \becomes{} \vec c_{(I)}
      \becomes{part.} \vec c_{I(I)}
      \becomes{sym.} \vec v_{I(I)}
      \quad \forall\,I.
      \end{equation}
\end{itemize}

\begin{figure}[ht]
\centering
\includegraphics[width=0.5\linewidth]{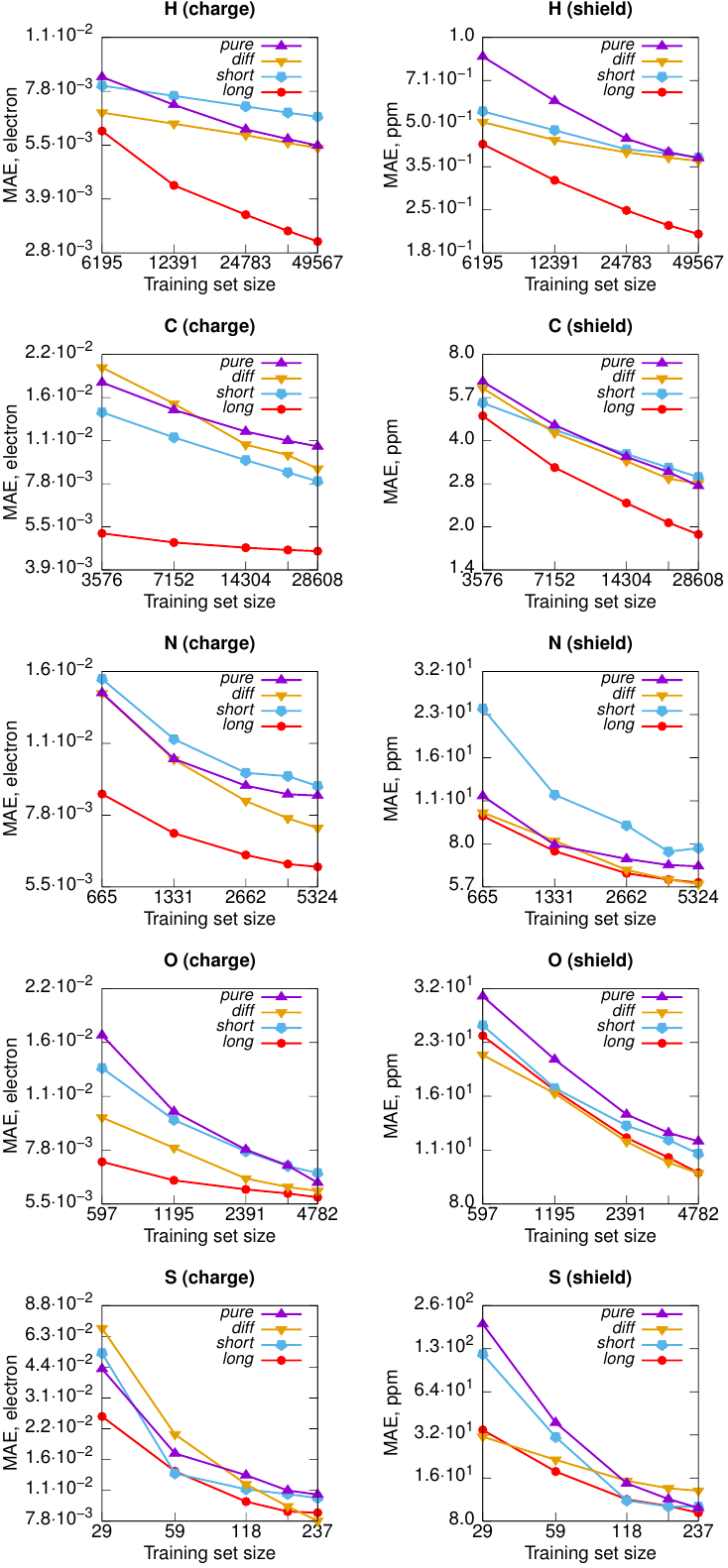}
\caption{
Learning curves of atomic charges and shielding constants
for the QM7 dataset.
The color code reflects the different models used to construct the \SPAHM[a] representation
from the rotationally-invariant vectors.
}
\label{fig:atom-models}
\end{figure}

The learning curves for the models are shown on Fig.~\ref{fig:atom-models}.
Overall, the \emph{long} model shows the best overall performance
and was selected as default to be used hereinafter.

\clearpage


\section{Generalization to open-shell systems}
\label{sec:alphabeta}

\begin{figure}[h]
\centering
\includegraphics[width=0.4\linewidth]{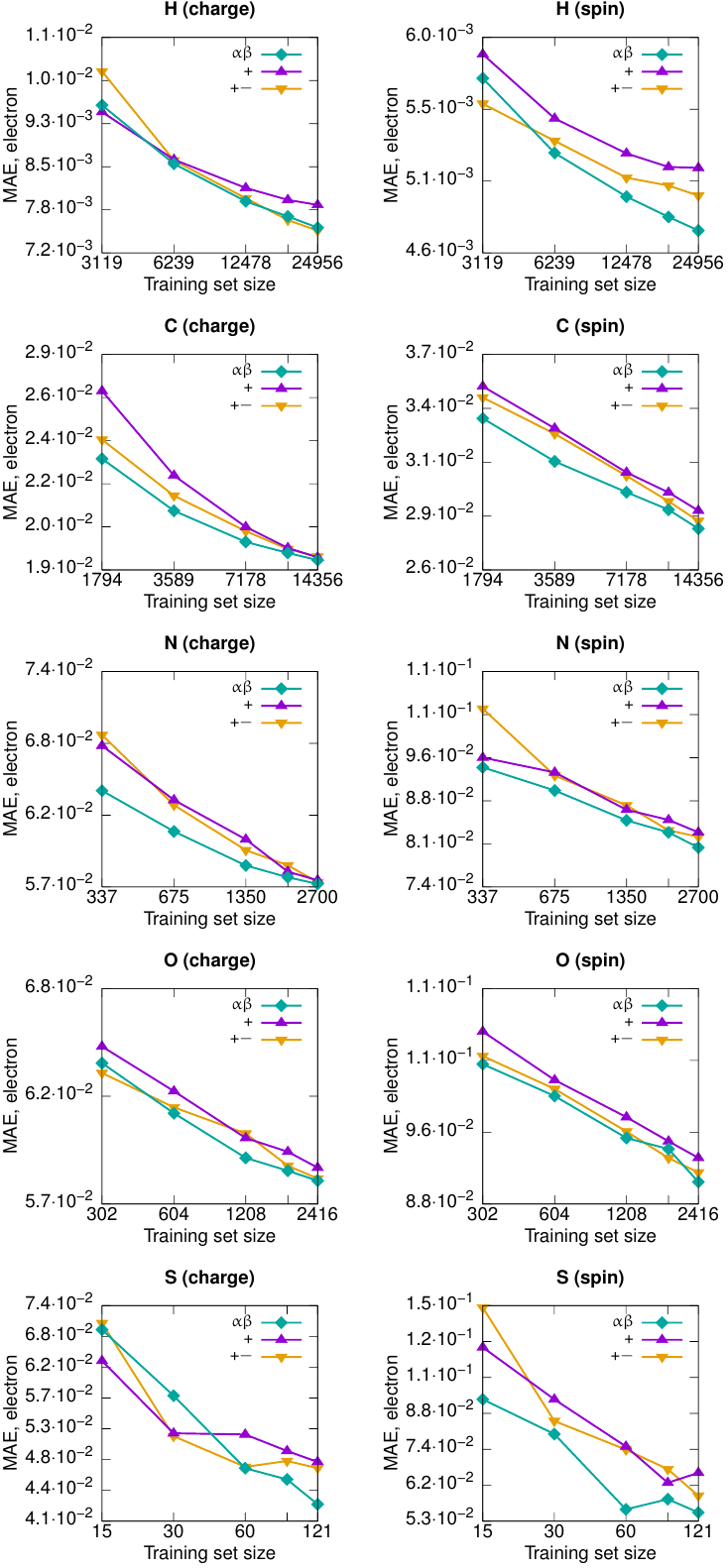}
\caption{
Learning curves of atomic charges and spins
for the QM7/2-RC dataset and the \SPAHM[b] representation.
The color code reflects the different models used to generalize the representation to open-shell systems.
}
\label{fig:alphabeta}
\end{figure}

We considered three ways to generalize the model to open-shell systems:
\begin{enumerate}[1)]
\item
concatenation of representation vectors $\vec x$
obtained from $\rho_\alpha$ and $\rho_\beta$ separately
(``$\alpha\beta$'');
\item
representation vector obtained from
$\rho = \rho_\alpha+\rho_\beta$,
the total electron density as in case of closed-shell systems~(``$+$'');
\item
concatenation of representation vectors
obtained from $\rho= \rho_\alpha+\rho_\beta$ and $\rho_m = \rho_\alpha-\rho_\beta$ separately
(``$+-$'').
\end{enumerate}
They were tested on the QM7/2-RC dataset with the \SPAHM[b] representation.
The results are shown on Fig.~\ref{fig:alphabeta}.
As expected, in most cases the ``$+$'' model,
having no information on the spin density, performed the worst,
whereas the ``$\alpha\beta$'' model showed the best results and was chosen as the default option.

\clearpage

\section{Basis set for the bond-density-based representation}

\subsection{Optimization}
\label{sec:basis}

The decomposition of the bond density onto a midbond-centered basis set
required optimization of a suitable basis.
First, we followed the procedure described in Ref.~\onlinecite{FBGC2020b}
used to optimize a basis to fit the on-top pair density.

\bigskip

For each bond of interest in a molecule, we search for the set of coefficients $\{ c_i \}$
that approximates the bond density in the least-squares sense,
\begin{equation}
\rho_{AB}(\vec r) \approx \sum_i c_i \phi_i(\vec r),
\qquad
\vec c = \vec S^{-1} \vec b,
\end{equation}
where $\vec S$ is the overlap matrix,
$b_i = \braket{\rho_{AB}|\phi_i}$, and
the decomposition error is
\begin{equation}
\mathcal E=
\int \Big(\rho_{AB}(\vec r) - \sum_i c_i \phi_i(\vec r) \Big)^2 \de^3 \vec r =
\braket{\rho_{AB} | \rho_{AB}} - \vec b\transp \vec S^{-1} \vec b.
\end{equation}
Thus,
to optimize the exponents, we minimize the sum of decomposition errors $\mathcal E$
for the molecules chosen for the bond of interest.
The exponents $\{\alpha_\mu\}$ for all the angular momenta are optimized simultaneously.
The exponents are parameterized as $\alpha_\mu = \exp(p_\mu)$,
and the first derivatives of the loss functions~$\mathcal E$ with respect to the exponents are computed as follows,
\begin{equation}
\pdd{\mathcal E}{\alpha_\mu} =
\vec c\transp \left( \pdd{\vec S}{\alpha_\mu}\, \vec c - 2\,\pdd{\vec b}{\alpha_\mu} \right),
\end{equation}
with the overlap integrals and their derivatives taken numerically.

All the bonds were treated separately.
For each bond (or atom pair) presented in the QM7 and APS datasets
we chose representative molecules containing it
(\eg, \ce{H2} for \ce{H}--\ce{H};
\ce{C2H2}, \ce{C2H4}, and \ce{C2H6} for \ce{C}--\ce{C};
\ce{H2O} and \ce{H2O2} for \ce{H}--\ce{O}),
and the sum of the molecular decomposition errors was minimized.
The maximum angular momentum $\ell_\mathrm{max}$ and the number of functions $n_\ell$ for each $\ell$
were gradually increased and optimized on each step,
until addition of further radial functions or
angular momenta did not provide any significant decrease of error.
The optimized exponents are available separately in \texttt{Q-stack}
(\url{https://github.com/lcmd-epfl/Q-stack}).

\bigskip

However, for some of the bonds the fitting errors were huge (up to 20\%)
due to the fact that largest fraction of the bond density is still localized on participating nuclei,
thus the fine-tuning of the fitting basis could not improve much.
This could be solved with adding a single Gaussian centered in the midbond as a weight function.
Our tests showed that, however the fitting error significantly decreased,
the quality of learning was almost the same.

\begin{figure}[ht]
\centering
\includegraphics[width=0.5\linewidth]{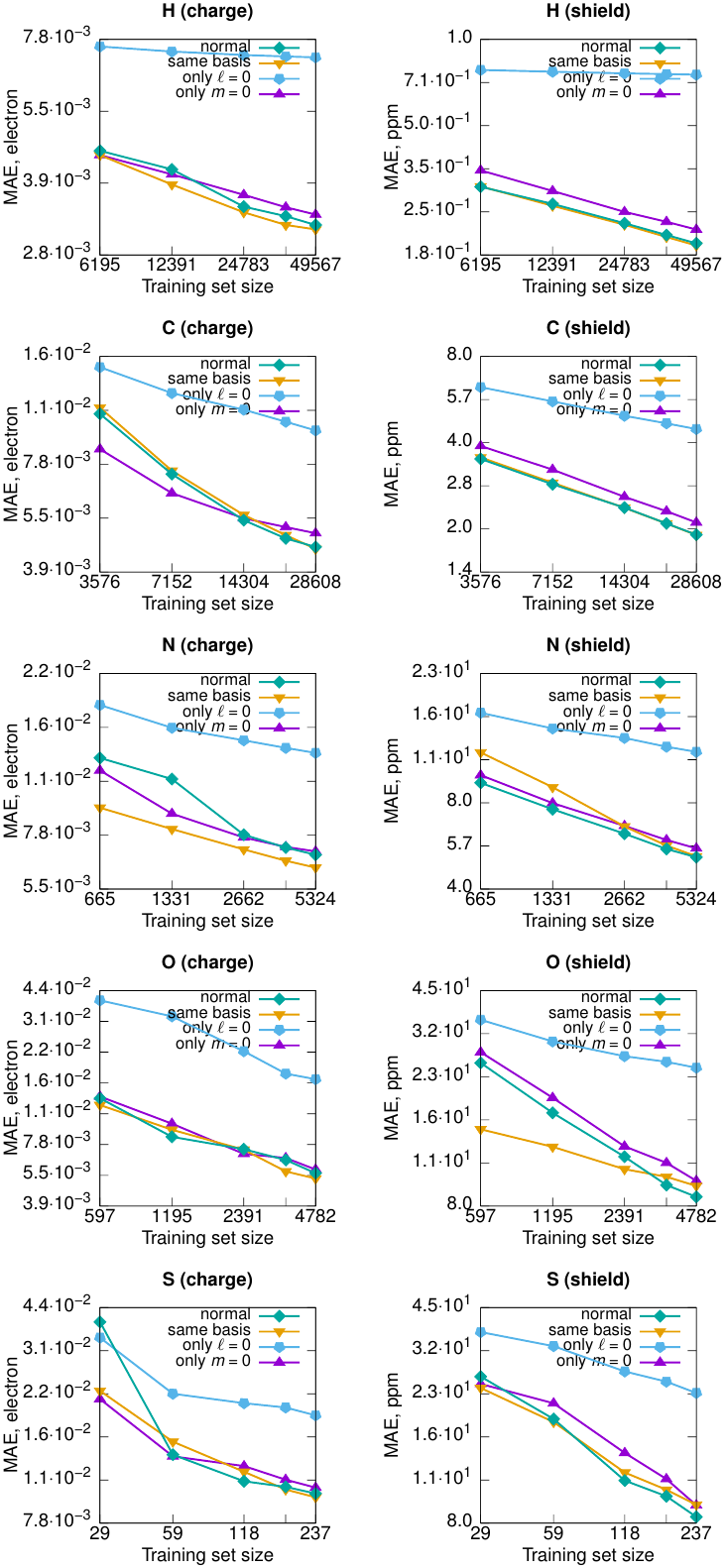}
\caption{
Learning curves of atomic charges and shielding constants
for the QM7 dataset.
The color code reflects the different basis sets used to generate the \SPAHM[b] representations:
``normal'':        fully-optimized basis for each bond;
``same basis'':    the same (\ce{C}--\ce{C}) basis for every bond;
``only $\ell=0$'': optimized basis with $s$-orbitals only;
``only $m=0$'':    optimized basis with $m\neq0$ orbitals excluded.
}
\label{fig:bond-basis}
\end{figure}

On Fig.~\ref{fig:bond-basis} we compare the performance of \SPAHM[b] computed
using the fully-optimized basis for each bond (``normal'')
and using the same (\ce{C}--\ce{C}) basis for every bond (``same basis'').
It is clear that the representation quality does not depend on the exponents of the basis
thus their optimization can be omitted.
(the role of angular momenta is discussed in Sec.~\ref{sec:simplebond}).

\clearpage


\subsection{Simplified models}
\label{sec:simplebond}

We also tested two approaches to simplify the bond-based representation,
which reduce the effort for both the two-electron integral evaluation and vector symmetrization.

\smallskip

The first one is to use only the $s$-orbitals.
The learning curves for the QM7 dataset
for the representation based on the fully-optimized basis truncated to the functions with $\ell=0$
are shown on Fig.~\ref{fig:bond-basis}.
Its peformance is significantly deteriorated and it is clear that higher angular momenta are necessary.

\smallskip

Another option is to use the orbitals with $m=0$, \ie, symmetric with respect to rotation around the bond.
Then Eq.~\ref{eq:bond-kernel-full} is simplified to
\begin{equation}
\kovlp{AB,XY}
=
\sum_{\ell \ell'}  \sum_{\substack{n_1n'_1 \\ n_2n'_2 }}
\underbrace{c_{n_1 \ell 0} c_{ n_1' \ell' 0} }_{u^{AB}_{p}}
\underbrace{
A_{n_1  n_2 }^{\ell}
A_{n_1' n_2'}^{\ell'}
}_{M_{pq}}
\underbrace{c_{ n_2 \ell 0} c_{n_2' \ell' 0 } }_{u^{XY}_{q}}.
\end{equation}
In the current implementation,
the bond density is first projected onto the DF basis set
and then rotated so the bond is aligned with the $z$-axis
and the DF coefficients are transformed accordingly.
This is why in our tests the density is fitted with the ``full'' basis set
and only the final representation is truncated to have only products of cylindrically-symmetric orbitals.

\begin{figure}[ht]
\centering
\includegraphics[width=0.46\linewidth]{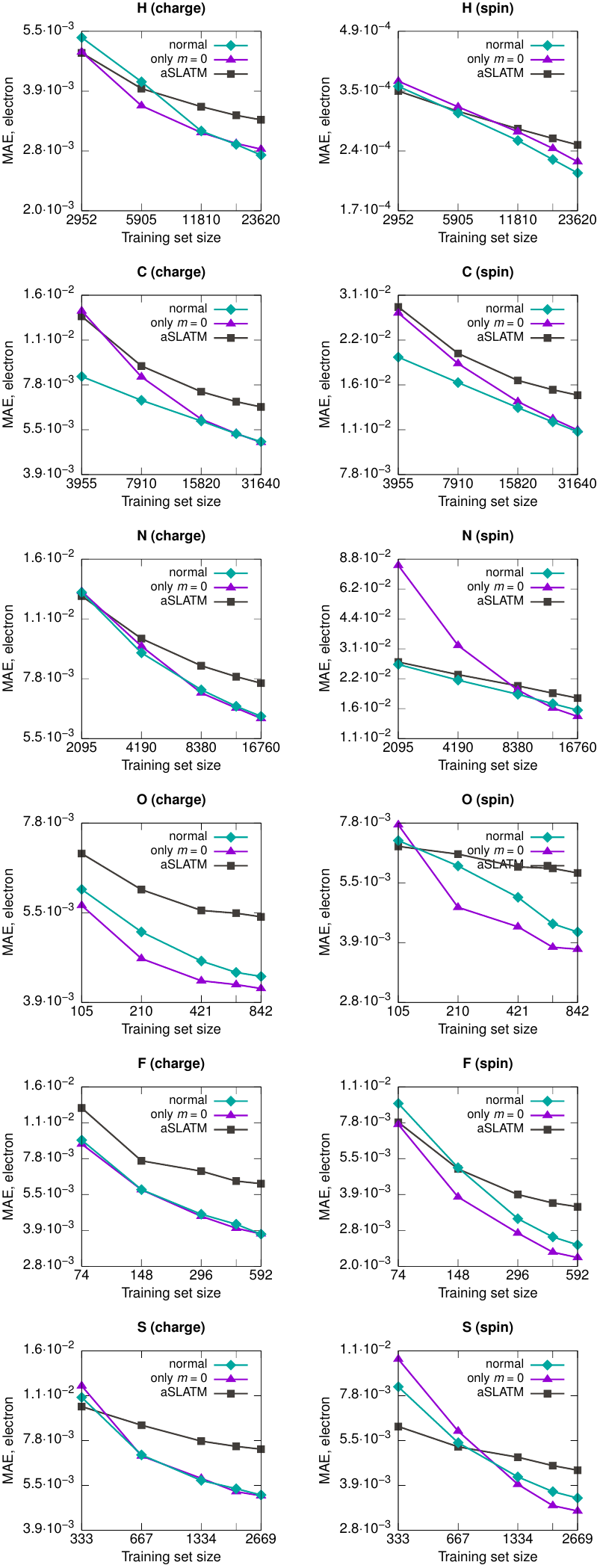}
\caption{
Learning curves of atomic charges and spins
for the APS-RC dataset.
The color code reflects the different basis sets used to generate the \SPAHM[b] representations:
``normal'': fully-optimized basis for each bond;
``only $m=0$'': optimized basis with $m\neq0$ orbitals excluded.
Learning curves for SLATM are given for comparison.
}
\label{fig:azoswitch-m0}
\end{figure}

The learning curves for QM7 and for APS-RC
comparing the truncated representation with the full one
are shown on Fig.~\ref{fig:bond-basis} and Fig.~\ref{fig:azoswitch-m0}, respectively.
For QM7, the truncated representation yields the same or slightly worse performance,
whereas for a more challenging APS-RC it even improves the learning in some cases.

\smallskip

While functions with $\ell=0$ are not sufficient to construct a good representation,
the representation built from $m=0$ only
performs very well on simple organic molecules
and at least in the case of the APS-RC dataset
the part of the density that seems to be orthogonal to the aromatic ring
is well enough captured by \eg $d_{z^2}$-orbital.
This simplification of \SPAHM[b] is promising in terms of both
performance and potential optimizations and should be studied further.

\clearpage

\section{Effect of the Hamiltonian}
\label{sec:potentials}

\begin{figure}[h]
\centering
\includegraphics[width=1.0\linewidth]{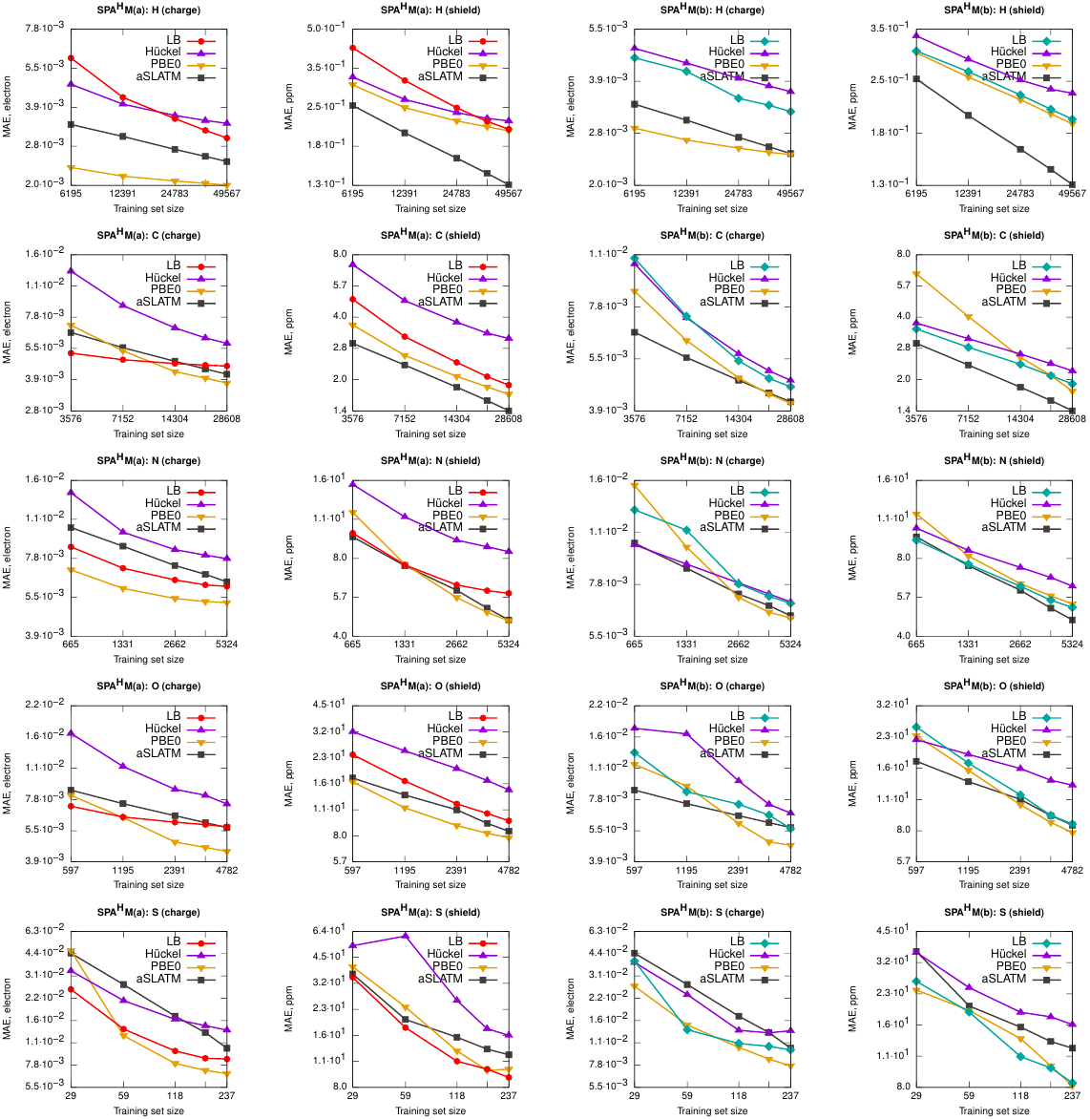}
\caption{
Learning curves of atomic charges and shielding constants
for the QM7 dataset.
The color code reflects the different Hamiltonians used to generate the \SPAHM[a,b] representations.
}
\label{fig:potentials}
\end{figure}

We compared the \SPAHM[a,b] representations built upon the density matrices obtained from
the H\"uckel guess\cite{H1963,L2019}, the LB\cite{LB2020} guess (default),
and a converged PBE0\cite{AB1999} computation.
The learning curves are shown on Fig.~\ref{fig:potentials}.
As expected, the worst approximation, the H\"uckel guess, gives the worst regression results.
In contrast to the eigenvalue \SPAHM,\cite{FBC2022},
the converged density makes the best representation,
sometimes overperforming SLATM, which opens the way to improvement of \SPAHM[a,b]
through improvement of the underlying guess Hamiltonian.

\clearpage

\section{Comparison with the KDFA representation}
\label{sec:mr2021}

Recently the kernel density functional approximation\cite{MR2021} (KDFA) was proposed,
similar in construction to our \SPAHM[a] model.

In KDFA, the representation vector for an atom
is also built from the density-fitting coefficients of the functions centered on its nucleus.
Instead of the coefficients themselves, rotationally-invariant sums $\sum_m |c_{nlm}|^2$ are used.
This could be seen as a simplification of Eq.~\ref{eq:atom-kernel-full}
with a combination of Kronecker deltas instead of $M_{pq}$,
omitting the cross-products of different radial basis functions,
\begin{equation}
K^\mathrm{KDFA}_{A,B} =
\sum_{\substack{\ell \\ n_1n'_1 \\ n_2n'_2 }}
\underbrace{
\left(\sum_m    c_{n_1\ell m}^A   c_{n_2\ell m}^A\right)
}_{u^A_p}
\underbrace{
\vphantom{\Bigg|}  \delta_{n_1 n_2} \delta_{n'_1 n'_2} \delta_{n_1 n'_1}
}_{M_{pq}}
\underbrace{
\left(\sum_{m} c_{n'_1\ell m}^B c_{n'_2\ell m}^B\right)
}_{u^B_q}
=
\sum_{n \ell}
\underbrace{
\left(\sum_m | c_{n\ell m}^A |^2\right)
}_{v^A_p}
\underbrace{
\left(\sum_{m} | c_{n\ell m}^B |^2\right)
}_{v^B_q}.
\end{equation}

\begin{figure}[ht]
\centering
\includegraphics[width=0.5\linewidth]{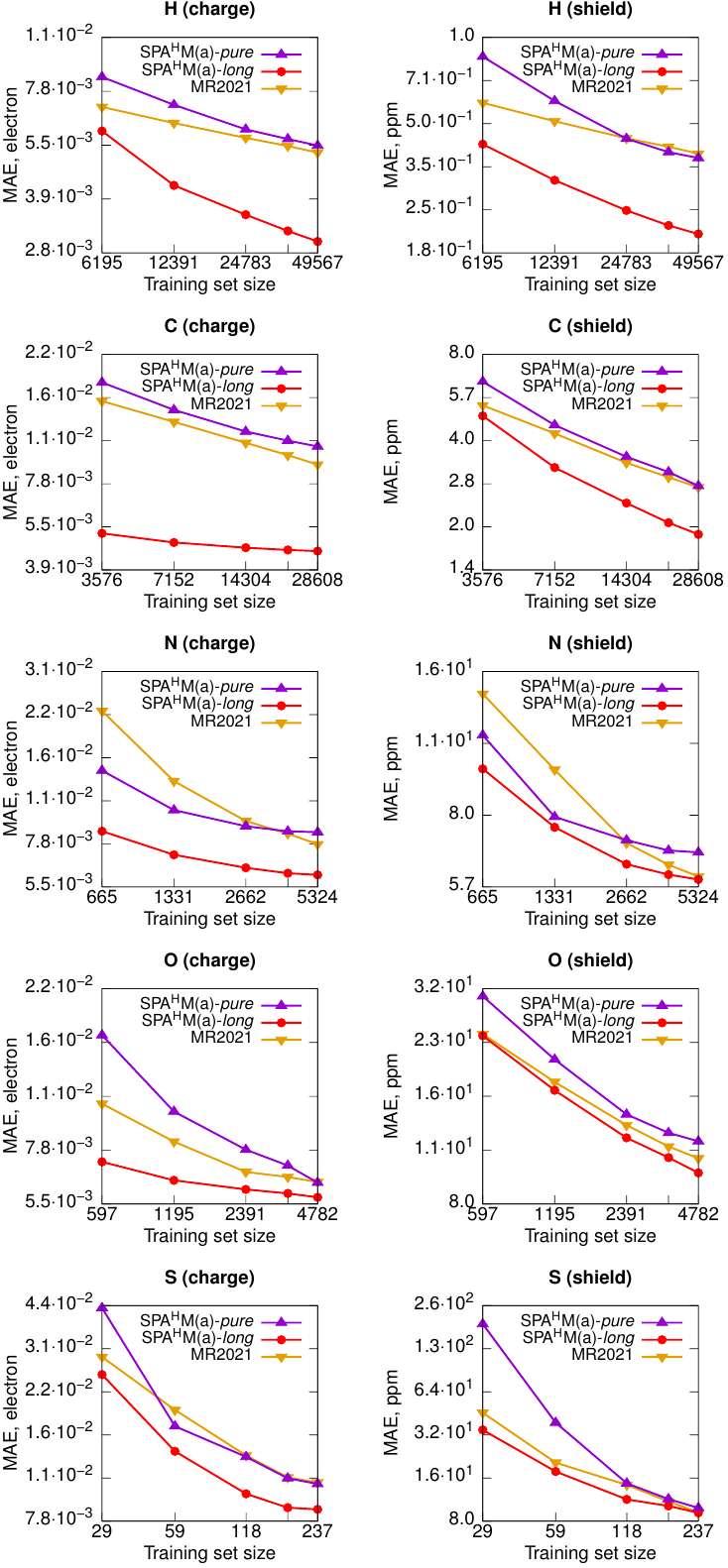}
\caption{
Learning curves of atomic charges and shielding constants
for the QM7 dataset.
The color code reflects the different representations.
``MR2021'' stands for the KDFA\cite{MR2021} representation.
}
\label{fig:mr2021}
\end{figure}

The learning curves comparing the performance of the KDFA representation
with our \emph{pure} and \emph{long} models (see Sec.~\ref{sec:models})
are shown of Fig.~\ref{fig:mr2021}.
Overall, the performance of the KDFA representation is close to the \emph{pure} model.
However, the \emph{long} model is consistently better,
presumably due to inclusion of ``long-range'' contributions to the atomic density.

\section*{References}

\bibliography{spahm+.bib}